\documentclass[fleqn,tradiabstract]{aa}
\usepackage{natbib}
\bibpunct{(}{)}{;}{a}{}{,}
\usepackage{graphicx}
\usepackage{amsmath}
\usepackage{amsfonts}
\usepackage{amssymb}

\newlength{\zero}
\newcommand{\mb}[1]{\mbox{\boldmath $#1$}}
\newcommand{\half}{{1\over2}}

\def\la{\mathrel{\hbox{\rlap{\hbox{\lower4pt\hbox{$\sim$}}}\hbox{$<$}}}}
\def\ga{\mathrel{\hbox{\rlap{\hbox{\lower4pt\hbox{$\sim$}}}\hbox{$>$}}}}
\def\ie{{i.e.}\,}
\def\eg{{e.g.}\,}

\begin{document}

\title{A new general relativistic magnetohydrodynamics code for
       dynamical spacetimes}

\author{Pablo Cerd\'a-Dur\'an \inst{1}
  \and Jos\'e A. Font \inst{2}
  \and Luis Ant\'on \inst{2}
  \and Ewald M\"uller \inst{1}}

\offprints{Pablo Cerd\'a-Dur\'an, \\ \email{cerda@mpa-garching.mpg.de}}

\institute{Max-Planck-Institut f\"ur Astrophysik,
  Karl-Schwarzschild-Str.~1, 85741 Garching, Germany
  \and
  Departamento de Astronom\'{\i}a y Astrof\'{\i}sica,
  Universidad de Valencia, 46100 Burjassot (Valencia), Spain}



\abstract{ We present a new numerical code that solves the general
relativistic magneto-hydrodynamical (GRMHD) equations coupled to the
Einstein equations for the evolution of a dynamical spacetime within a 
conformally-flat approximation.  This
code has been developed with the main objective of studying
astrophysical scenarios in which both, high magnetic fields and strong
gravitational fields appear, such as the magneto-rotational collapse
of stellar cores, the collapsar model of GRBs, and the evolution of
neutron stars.  The code is based on an existing and thoroughly tested
purely hydrodynamical code and on its extension to accommodate weakly
magnetized fluids (passive magnetic-field approximation). These codes
have been applied in the past to simulate the aforementioned
scenarios with increasing levels of sophistication in the input
physics. The numerical code we present here is based on
high-resolution shock-capturing schemes to solve the GRMHD equations,
which are cast in first-order, flux-conservative hyperbolic form,
together with the flux constraint transport method to ensure the
solenoidal condition of the magnetic field. Since the astrophysical
applications envisaged do not deviate significantly from spherical symmetry,
the conformal flatness condition approximation is used for the
formulation of the Einstein equations; this has repeatedly shown to
yield very good agreement with full general relativistic simulations
of core-collapse supernovae and the evolution of isolated neutron stars. In
addition, the code can handle several equations of state, from simple
analytical expressions to microphysical tabulated ones. In this paper
we present stringent tests of our new GRMHD numerical code, which show
its ability to handle all aspects appearing in the astrophysical
scenarios for which the code is intended, namely relativistic shocks,
highly magnetized fluids, and equilibrium configurations of magnetized
neutron stars. As an application, magneto-rotational core-collapse
simulations of a realistic progenitor are presented and the
results compared with our previous findings in the passive magnetic-field
approximation.}


\authorrunning{Cerd\'a-Dur\'an et al.}
\titlerunning{A new general relativistic magnetohydrodynamics code}

\maketitle

\section{Introduction}
The collapse of rotating stellar cores and the merging of compact
binaries (either neutron star-neutron star or neutron star-black hole
binaries) are two of the most important astrophysical scenarios
involving compact objects, whose modeling requires the study of the
dynamical evolution of a magnetized fluid in general relativity. In
the case of the rotational collapse of massive stellar cores, the
magnetic field is thought to grow through the extraction of energy
from the differential rotation generated during collapse
\citep{meier76}. This idea is supported by the observational fact that
some neutron stars (magnetars) possess extremely large magnetic fields
($10^{14}-10^{15}\,$Gauss), as inferred from studies of anomalous
X-ray pulsars and soft gamma-ray repeaters \citep{kouveliotou98}.
Furthermore, the class of soft-long gamma-ray bursts (GRB) are
probably the result of jets formed in a rotational core-collapse event
leading to a black hole, according to the collapsar scenario
\citep{woosley1993}. In this case, the magnetic field most likely
plays a crucial role in the formation and collimation of the jet. This
scenario is supported by the existing correlation of some long GRB
with core-collapse supernova events \citep[see][and references
therein]{kelly2007}. In the case of the mergers of two neutron stars,
believed to be the standard mechanism to account for hard-short GRBs,
it was suggested \citep{price2006} that the strong shear between
the two neutron stars could generate strong magnetic fields, too.

A considerable effort has been made to develop special
relativistic magneto-hydrodynamics (SRMHD) codes (see \eg
\cite{LRmarti}, \cite{ibanez06} and references therein). 
Most works have considered the
case of {\it ideal} MHD where the fluid is assumed to be a perfect
conductor. In this case, the resulting system of equations is
simplified significantly, and can be solved by numerical codes
designed specifically for hyperbolic systems.  These codes include the
use of Godunov-type schemes \citep{komissarov99}, numerical techniques
to keep the magnetic field divergence-free \citep[see][and references
therein]{toth00}, and efficient recovery schemes to derive
primitive quantities from the conserved ones \citep[see][and
references therein]{noble2006}. There is also a
major activity in the development of codes capable of simulating
magnetized astrophysical flows in general relativity. These codes
integrate the ideal GRMHD equations for fixed background spacetimes
using high-order conservative schemes based on either approximate or
full wave-decomposition Riemann solvers \citep{Gammie03,komissarov05,
cosmos++,anton06,echo,wham}.  Resistive MHD flows were considered
by \cite{komissarov07}, and nonconservative GRMHD schemes and schemes
relying on artificial viscosity were used
by \cite{devilliers} and \cite{cosmos++}. Most codes have been applied to study
disk accretion onto black holes and jet formation, but since the
self-gravity of the fluid was not taken into account, these codes
cannot simulate consistently the formation of the black hole and the
evolution of the surrounding disk or torus.

Only very recently, GRMHD codes are able to follow the evolution of a
dynamical spacetime. The codes of \cite{Duez05}, \cite{shibata05}, 
and \cite{giacomazzo07}
are based on the BSSN formulation of the Einstein equations for the
spacetime \citep{NOK,SN95,BS99}, high resolution shock-capturing schemes
for the GRMHD equations involving approximate Riemann solvers such as
HLL or high-order central schemes, and on the constraint transport
scheme for the magnetic field.  In the code of \cite{anderson08}, the
Einstein equations are cast in first-order symmetric hyperbolic
form, and are solved using the generalized harmonic
decomposition. While the code relies on the same type of GR
hydrodynamics solvers as previously developed codes, it guarantees
a divergence-free magnetic field by means of either projection methods
or hyperbolic divergence cleaning.  Both methods are easier to
implement for the non-structured AMR grids employed in this code than
the constrained transport method. All four codes rely on Cartesian
coordinates for three dimensional simulations. The codes of
\cite{Duez05} and \cite{shibata05} also provide the possibility to
impose axisymmetry by means of the cartoon method for the spacetime
evolution and the use of cylindrical coordinates for the GRMHD
equations.  The equations of state (EOS) implemented in these codes
consist of simple analytic expressions: polytropic EOS, ideal gas or
hybrid EOS (see Sect. \ref{sec:eos}).  One of the
codes was extended to handle a tabulated microphysical EOS
\citep{shibata07}.

We present a new axisymmetric numerical code, cable of
handling ideal MHD flows in dynamical spacetimes in general relativity,
and designed particularly to investigate gravitational core collapse.
We use similar numerical schemes as in most of the other existing
GRMHD codes (HRSC schemes and constraint transport), but we follow a
simpler approach for the spacetime evolution.

The new code is based on the hydrodynamics code described in
\citet{dimmelmeier_02_a, dimmelmeier_02_b}, and on its extensions
discussed in \cite{cerda05}, \cite{nfnr}, and \cite{cerda07}.  The
Maxwell equations are already incorporated in the codes of
\cite{nfnr} and \cite{cerda07}, but only in the passive 
magnetic-field approximation, \ie the contribution of the magnetic field to
the energy-momentum tensor is neglected, and therefore has no impact on the
dynamics. In the new code, we relax this assumption and incorporate
magnetic field effects on the spacetime dynamics and the self-gravity
of the fluid following the approach laid out in \cite{anton06}.  The
Einstein equations are formulated using the conformal flatness
condition (CFC hereafter). This approximate treatment of the metric
equations was first introduced by \cite{isenberg_78_a} and
\cite{wilson_96_a}, and was used to study rotational core
collapse \citep{dimmelmeier_02_a, dimmelmeier_02_b}, and binary
neutron stars~\citep{oechslin07}.  Simulations with a second
post-Newtonian extension of the CFC metric (named CFC+) showed small
quantitative differences in the dynamics and the gravitational
waveforms ($<1\%$) compared to the CFC metric, both for rotational
core collapse and for simulations of the evolution of single neutron
stars \citep{cerda05}. Direct comparisons of the CFC approach with full
general relativistic simulations were reported by
\cite{shibata_04_a} and \cite{ott07a,ott07b}, who found that the
differences in the collapse dynamics and the waveforms are minute
demonstrating the suitability of CFC for performing accurate core
collapse simulations.

The CFC approach has some advantages compared with the BSSN formulation:
(i) the Hamiltonian and momentum constraints of the Einstein equations
are automatically satisfied, and (ii) the time step is less
restrictive since it is determined by the largest fluid eigenvalue,
while in hyperbolic formulations (such as BSSN) the largest eigenvalue
is the speed of light. Consequently, the time steps are typically ten
times larger than those admitted in BSSN for the same grid for the
evolution of neutron stars, and even larger during core collapse.
However, there are also some disadvantages of the CFC approach.  It
neglects the gravitational wave content of the spacetime, \ie when
the gravitational wave back reaction is important (\eg in neutron-star 
mergers) the dynamics cannot be modeled accurately (\eg compare
the simulations of \cite{oechslin07} and \cite{shibata06d}).
Furthermore, to compute the gravitational waveforms one needs
to resort to the Einstein quadrupole formula, and, since the CFC metric
equations are elliptic, the parallelization of the code for a
large number of processors is more difficult than in hyperbolic formulations
such as BSSN, but still possible.

Our code uses spherical polar coordinates and (presently) assumes
axisymmetry.  The most important advantage of these coordinates with
respect to Cartesian or cylindrical coordinates adopted by other
numerical codes, is that they are more readily adapted to the astrophysical
scenarios that we wish to study.  Furthermore, it allows us to easily
and properly cover the length-scales of core collapse ranging from the
radius of the initial iron core ($\sim 1000$\,km) down to the radius
of the neutron star ($\sim 10$\,km) by means of a logarithmically
spaced radial grid.  A disadvantage of our coordinate system concerns
its possible extension to 3D because of the coordinate singularities
at the center and at the axis. Moreover, as the azimuthal grid spacing
decreases quadratically towards the axis with increasing grid
resolution (for an equally-spaced angular grid), the Courant condition
for the time step can be rather restrictive in 3D simulations
(a possible solution to this issue can be found e.g. in \cite{Zink2008a}).

The code can handle various equations of state ranging from simple
analytical expressions (polytropes, ideal gas and hybrid EOS) to
tabulated microphysical EOS. General relativistic hydrodynamic core-collapse 
simulations using the tabulated EOS were performed by
\cite{ott07a} and \cite{dimmelmeier_07_a}, and magneto-hydrodynamic
simulations by \cite{cerda07} using the passive magnetic-field
approximation. These three studies also included a simplified
treatment of neutrino transport.

The rest of the paper is organized as follows.
Section \ref{sec:formulation} presents a brief overview of the
theoretical framework we use, namely the CFC equations and the GRMHD
equations in the $3+1$ formalism. Our numerical approach is discussed
in Sect.\,\ref{sec:numerical}. Tests of the numerical code are
presented in Sect.\,\ref{sec:tests} including a magneto-rotational
core collapse simulation, and the conclusions are given in
Sect.\,\ref{sec:conclusions}. Throughout the paper, we use a spacelike
metric signature $ (-, +, +, +) $, and units where $ c = G = 1 $. We
absorb the factor $1/\sqrt{4 \pi}$ appearing in the MHD equations in
the definition of the magnetic field $B^i$, \ie the units of the
magnetic field are $\sqrt{4 \pi}$~Gauss. Greek indices run from 0 to
3, Latin indices from 1 to 3, and we adopt the standard Einstein
summation convention.

\section{Physical approach}
\label{sec:formulation}
We adopt the $ 3 + 1 $ formalism of general relativity
\citep{lichnerowicz44} to foliate the spacetime into spacelike
hypersurfaces. In this approach, the line element reads
\begin{equation}
  ds^2 = - \alpha^2 \, dt^2 + \gamma_{ij} (dx^i + \beta^i \,dt)
  (dx^j + \beta^j \, dt),
\end{equation}
where $ \alpha $ is the lapse function, $ \beta^i $ is the shift
vector, and $ \gamma_{ij} $ is the spatial three-metric induced in
each hypersurface. Using the projection operator $ \perp^\mu_\nu $ and
the unit four-vector $ n^\mu $ normal to each hypersurface, it is
possible to build the quantities
\begin{eqnarray}
  E & \equiv & n^{\mu} n^{\nu} T_{\mu\nu} = \alpha^2 T^{00},
  \label{eq:tmn_projection_1}
  \\
  S_i & \equiv & - \perp^{\mu}_{i} n^{\nu} T_{\mu\nu} =
  - \frac{1}{\alpha} (T_{0i} - T_{ij} \beta^j),
  \label{eq:tmn_projection_2}
  \\
  S_{ij} & \equiv & \perp^{\mu}_i \perp^{\nu}_j T_{\mu\nu} = T_{ij},
  \label{eq:tmn_projection_3}
\end{eqnarray}%
which represent the total energy, the momenta, and the spatial
components of the energy-momentum tensor $T_{\mu \nu}$, respectively.

To solve the gravitational field equations we choose the ADM gauge in
which the three-metric can be decomposed as $ \gamma_{ij} = \phi^4
\hat{\gamma}_{ij} + h^\mathrm{TT}_{ij}$, where $ \phi $ is the
conformal factor, $\hat{\gamma}_{ij}$ is the flat three-metric, and $
h^\mathrm{TT}_{ij} $ is the transverse and traceless part of the
three-metric. We note that this gauge choice implies the maximal slicing
condition where the trace $ K $ of the extrinsic curvature tensor $
K_{ij} $ vanishes.

\subsection{The CFC approximation}
In our work, Einstein's field equations are formulated and solved using
the conformally flat condition (CFC hereafter), introduced by
\citet{isenberg_78_a} and first used in a dynamical context by
\citet{wilson_96_a}. In this approximation, the three-metric in the
ADM gauge is assumed to be conformally flat, $ \gamma_{ij} = \phi^4
\hat\gamma_{ij} $. We note that this approximation can also be realized
for other gauge choices such as the quasi-isotropic gauge or the Dirac
gauge, both supplemented by the maximal slicing condition. Under the
CFC assumption, the gravitational field equations can be written as a
system of five nonlinear elliptic equations,
\begin{eqnarray}
  \hat{\Delta} \phi & = & - 2 \pi \phi^5
  \left( E + \frac{K_{ij}K^{ij}}{16 \pi} \right),
  \label{eq:cfc1}
  \\
  \hat{\Delta} (\alpha \phi) & = & 2 \pi \alpha \phi^5
  \left( E + 2 S + \frac{7 K_{ij}K^{ij}}{16 \pi} \right),
  \label{eq:cfc2}
  \\
  \hat{\Delta} \beta^i & = & 16 \pi \alpha \phi^4 S^i +
  2 \phi^{10} K^{ij} \hat{\nabla}_j
  \left(\! \frac{\alpha}{\phi^6} \!\right) -
  \frac{1}{3} \hat{\nabla}^i \hat{\nabla}_k \beta^k,
  \label{eq:cfc3}
\end{eqnarray}
where $ \hat{\Delta} $ and $ \hat{\nabla} $ are the Laplace and nabla
operators associated with the flat three-metric, and
$ S \equiv \gamma^{ij} S_{ij} $.

\subsection{General relativistic magnetohydrodynamics}
The energy-momentum tensor of a magnetized perfect fluid can be
written as the sum of the fluid part and the electromagnetic field
part. In the so-called ideal MHD limit (where the fluid is a perfect
conductor of infinite conductivity), the latter can be expressed
solely in terms of the magnetic field $ b^\mu $ measured by a {\it
comoving} observer. In this case, the total energy-momentum tensor is
given by
\begin{equation}
  T^{\mu \nu} = (\rho h + b^2) \, u^\mu u^\nu +
  \left( P + \frac{b^2}{2} \right) g^{\mu \nu} - b^\mu b^\nu,
  \label{eq:tmunu_grmhd}
\end{equation}
where $ \rho $ is the rest-mass density, $ h = 1 + \epsilon + P / \rho
$ the relativistic enthalpy, $ \epsilon $ the specific internal
energy, $ P $ the fluid pressure, $ u^\mu $ the four-velocity of the
fluid, and $ b^2 = b^\mu b_\mu $. We define the magnetic pressure $
P_\mathrm{mag} = b^2 / 2 $ and the specific magnetic energy $
\epsilon_\mathrm{mag} = b^2 / (2 \rho) $, whose effect on the dynamics
is similar to that of the fluid pressure and the specific internal
energy of the fluid, respectively.  

For an {\it Eulerian} observer, $ u^\mu = n^\mu $, and in the ideal
MHD limit, the temporal component of the electric field vanishes, $
E^\mu = (0, - \varepsilon_{ijk} v^j B^k) $, where $\varepsilon_{ijk}$
is the permutation tensor and $B^k$ is the magnetic field. In this
case, Maxwell's equations reduce to the divergence-free condition and
the induction equation for the magnetic field,
\begin{equation}
  \hat{\nabla}_i B^{*\,i} = 0,
  \qquad
  \frac{\partial B^{*\,i}}{\partial t} =
  \hat{\nabla}_j (v^{*\,i} B^{*\,j} - v^{*\,j} B^{*\,i}),
\end{equation}
with $ B^{*\,i} \equiv \sqrt{\bar{\gamma}} B^i $ and $ v^{*\,i} \equiv
\alpha v^i - \beta^i $, where $ v^i $ is the fluid three-velocity as
measured by the Eulerian observer. The ratio of the determinants of
the three-metric and the flat three-metric is given by $ \bar{\gamma}
= \gamma / \hat{\gamma} $.

The evolution of a magnetized fluid is determined by the conservation
law of the energy-momentum, $ \nabla_\mu T^{\mu \nu} = 0 $, and by the
continuity equation, $ \nabla_\mu J^\mu = 0 $, for the rest-mass
current $ J^\mu = \rho u^\mu $. Following the procedure described by
\citet{anton06}, the conserved quantities are chosen in a way similar
to the purely hydrodynamic case presented by \citet{banyuls_97_a}:
\begin{eqnarray}
  D & = & \rho W,
  \label{eq:def:d}\\
  S_i & = & (\rho h + b^2) W^2 v_i - \alpha b_i b^0,
  \label{eq:def:si}\\
  \tau & = & (\rho h + b^2) W^2 - \left( P + \frac{b^2}{2} \right) -
  \alpha^2 (b^0)^2 - D,
  \label{eq:def:tau}	
\end{eqnarray}%
where $ W = \alpha u^0 $ is the Lorentz factor. With this choice, the
system of conservation equations for the fluid, and the induction
equation for the magnetic field can be cast as a first-order,
flux-conservative, hyperbolic system,
\begin{equation}
  \frac{1}{\sqrt{- g}} \left[
  \frac{\partial \sqrt{\gamma} \mb{U}}{\partial t} +
  \frac{\partial \sqrt{- g} \mb{F}^i}{\partial x^i} \right] = \mb{S},
  \label{eq:hydro_conservation_equation}
\end{equation}
with the state vector, the flux vector, and the source vector given by
\begin{eqnarray}
  \mb{U} & = & [D, S_j, \tau, B^k],
  \label{eq:state_vector}
  \\
  \mb{F}^i & = & \left[
  D \hat{v}^i, S_j \hat{v}^i + \delta^i_j \left( P + \frac{b^2}{2} \right) -
  \frac{b_j B^i}{W}, \right.
  \nonumber
  \\
  & & \left. \tau \hat{v}^i + \left( P + \frac{b^2}{2} \right) v^i -
  \alpha \frac{b^0 B^i}{W}, \hat{v}^i B^k - \hat{v}^k B^i \right],
  \label{eq:flux_vector}
  \\
  \mb{S} & = & \left[ 0, \frac{1}{2} T^{\mu \nu}
  \frac{\partial g_{\mu \nu}}{\partial x^j},
  \alpha \! \left( \! T^{\mu 0} \frac{\partial \ln \alpha}{\partial x^\mu} -
  T^{\mu \nu} {\it \Gamma}^0_{\mu \nu} \! \right), 0^k \! \right],
  \label{eq:source_vector}
\end{eqnarray}%
where $ \delta^i_j $ is the Kronecker delta, and $ \Gamma^\mu_{\mu
\lambda}$ are the Christoffel symbols associated with the four-metric.
We note that the above definitions contain components of the magnetic
field measured by both a comoving observer and an Eulerian observer.
The two are related by
\begin{equation}
  b^0 = \frac{W B^i v_i}{\alpha},
  \qquad
  b^i = \frac{B^i + \alpha b^0 u^i}{W}.
\end{equation}
The hyperbolic structure of Eq.\,(\ref{eq:hydro_conservation_equation}) 
and the associated spectral decomposition (into eigenvalues and
eigenvectors) of the flux-vector Jacobians are given in \citet{anton06}. 
This information is required to numerically solve the system of 
equations using the class of high-resolution shock-capturing schemes 
that we have implemented in our code.

\subsection{Equation of state}
\label{sec:eos}
The new numerical code can handle a variety of equations of state
including a polytropic EOS, an ideal gas EOS, a hybrid EOS, and a
tabulated microphysical EOS.

\subsubsection{Hybrid EOS}
\label{sec:hybrid_eos}
The hybrid EOS \citep{janka_93_a} is a simplified analytical equation
of state used in core collapse simulations. The pressure is given by a
polytropic part, $ P_\mathrm{p} = K \rho^\gamma $, with $ K = 4.897
\times 10^{14} $ (in cgs units), plus a thermal part, $ P_\mathrm{th}
= \rho \epsilon_\mathrm{th} (\gamma_\mathrm{th} - 1) $, where the
specific thermal energy, $ \epsilon_\mathrm{th} = \epsilon -
\epsilon_\mathrm{p} $, and $ \gamma_\mathrm{th} = 1.5 $. The thermal
contribution takes into account the increase of the thermal energy due to
shock heating. When $ \rho $ exceeds the nuclear saturation density, $
\rho_\mathrm{nuc} = 2.0 \times 10^{14} \mathrm{\ g\ cm}^{-3} $, the
value of $ \gamma $ is raised to $ \gamma_2 = 2.5 $, and $ K $ is
adjusted accordingly to guarantee the continuity of $ P $ and $
\epsilon $. Due to this stiffening of the EOS the core undergoes a
so-called pressure-supported bounce. More details about the hybrid EOS
can be found, \eg in \citet{dimmelmeier_02_a}.

\subsubsection{Microphysical EOS}
\label{sec:SHEN}
We further employ the tabulated non-zero temperature nuclear EOS by
\citet{shen_98_a} in the variant of \citet{marek_05_a}, which includes
baryonic, electronic, and photonic pressure components. It specifies the
fluid pressure $ P $ (and additional thermodynamic quantities) as a
function of $ \rho $, the temperature $ T $, and the electron fraction
$ Y_e $.  Whenever it is necessary in the code to compute the pressure
as a function of the specific internal energy $ \epsilon $ instead of
the temperature $ T $, we iterate the corresponding value of $ T $
with a Newton--Raphson scheme.

\section{Numerical methods}
\label{sec:numerical}
Since our new numerical code is based on a previous purely
hydrodynamic code \citep{dimmelmeier_02_a, dimmelmeier_02_b} and on
its extension to the passive magnetic-field approximation
\citep{nfnr,cerda07}, we describe here in detail only those numerical 
techniques that represent improvements over their
predecessors, and provide only concise information about the numerical
schemes already described and tested elsewhere.

The code solves the coupled time evolution of the equations governing
the dynamics of the spacetime, the fluid, and the magnetic field in
general relativity. The equations are implemented in the code using
spherical polar coordinates $ \{ t, r, \theta, \varphi \} $. We assume
axisymmetry and equatorial plane symmetry.

\subsection{Metric solver}
The CFC metric equations, Eqs.\,(\ref{eq:cfc1}-\ref{eq:cfc2}), are
five nonlinear elliptic coupled Poisson-like equations, which can be
written in compact form as $ \hat{\Delta} \mb{u} (\mb{x}) = \mb{f}
(\mb{x}; \mb{u} (\mb{x}))$, where $ \mb{u} = u^k = (\phi, \alpha \phi,
\beta^j) $, and $ \mb{f} = f^k $ is the source vector. These five
scalar equations are coupled via the source vector, which depends on
the components of $ \mb{u} $. We use a fix-point iteration scheme in
combination with a linear Poisson solver to solve these equations (for
further details see \citet{cerda05} and \citet{dimmelmeier_02_a}).

Since the CFC equations are written in terms of the energy-momentum
tensor, the contribution of the energy-momentum of the magnetic field
is automatically incorporated in the computations. The main difference
with respect to the non-magnetized case arises from the fact that the
magnetic-field contribution does not necessarily have compact support.
However, the magnetic field far from the fluid should decay at
least as a dipole ($\sim 1/r^3$), \ie its contribution can be
computed correctly by integrating the CFC equations in a sufficiently
large volume. We checked that in all cases considered here
the contribution of the outer magnetic field to the energy-momentum
tensor is too small to affect the CFC metric.  Hence, we consider 
the contribution of the magnetic field only up to a radius $\sim 20\%$
larger than the radial extent of the fluid. The contribution of the
inner magnetic field is, however, in some case sufficiently large to
modify the metric (\eg for the magnetized neutron-star equilibria),
and therefore cannot be neglected in the CFC equations.

\subsection{Riemann solver}
For the evolution of the matter fields we utilize a HRSC scheme to
integrate the subset of equations in the
system of Eq.~(\ref{eq:hydro_conservation_equation}) that corresponds to the
hydrodynamic variables ($ D $, $ S_i $,$ \tau $).  HRSC schemes ensure
the numerical conservation of physically conserved quantities and a
correct treatment of discontinuities such as shocks \citep[see
\eg][for a review and references therein]{font_03_a}. We
implemented various cell-reconstruction procedures that are accurate to
either second-order or third-order in space, namely minmod, MC, and
PHM \citep[see][for definitions]{toro99}. The time update of the state
vector $ \mb{U} $ relies on the method of lines in combination with a
second-order accurate Runge--Kutta scheme. The numerical fluxes at
cell interfaces are obtained using either the HLL single-state solver
of \citet{harten83} or the symmetric scheme of \citet{kt00} (KT
hereafter). Both solvers yield results with an accuracy comparable to
Riemann solvers exploiting the full characteristic information, as
demonstrated for hydrodynamic special relativistic \citep{lucas04} and
general relativistic flows in dynamical spacetimes
\citep{shibata_font05}. Tests of both solvers in GRMHD were
reported by \citet{anton06}.

\subsection{Constrained transport scheme}
The evolution of the magnetic field needs to be performed differently
from the rest of the conservation equations because the physical meaning
of the corresponding conservation equation is different. Although the
induction equation can be written in a flux-conservative form, a
supplementary condition for the magnetic field (the divergence
constraint, or the conservation of the magnetic flux) has to be
fulfilled during the whole evolution.  Among the numerical schemes
that satisfy this condition \citep[see][for a review]{toth00}, the
constrained transport (CT) scheme \citep{evans88} was proven to be
adequate for performing accurate simulations of magnetized flows. Our
particular implementation of the CT scheme is adapted to the
spherical polar coordinates used in the code, and uses cell
interface-centered poloidal and (because of the assumption of
axisymmetry) cell-centered toroidal magnetic-field components
\citep[see Sect.\,3.2.1 in][for details]{nfnr}. The induction 
equation is discretized in the same way as for the fluid equations.

CT schemes preserve the magnetic flux during the evolution of a
magnetized flow, but do not impose the divergence constraint on the
initial magnetic field. Hence, one also has to provide initial data
that fulfill this constraint in order for the CT method to work
properly. This can be ensured by computing the staggered magnetic
field from the vector potential \citep[see Eqs.\,(28) and (29) in][for
details]{nfnr}.

Finally, one has to consider the computation of cell-centered values
of the (poloidal) magnetic field, which are required in the source
terms and for the reconstruction of the magnetic field tangential to the
cell interfaces, that enter the evaluation of the numerical flux. Here,
we depart from the scheme described by \cite{anton06}, who computed
cell-centered magnetic field components assuming that the
corresponding magnetic flux at the cell center is given by the average
of the magnetic flux at the cell interfaces.  Using this prescription,
the cell-centered magnetic pressure differs from that at the
interface, even in the case of a homogeneous magnetic field. Instead we
use 
\begin{eqnarray}
B^{*r}_{i\ j} 
   &=& - \cos{\theta_{i\ j}}\
       \frac{\cos \theta_{j+\half} - \cos \theta_{j-\half} }
       { \sin^2 \theta_{j+\half} - \sin^2 \theta_{j-\half}}
       (B^{*r}_{i+\half\ j} + B^{*r}_{i-\half \ j}) \nonumber\\
B^{*\theta}_{i\ j} 
   &=& \frac{\sin \theta_{i\ j}}{2}
       \left(  \frac{B^{*\theta}_{i\ j+\half}}{\sin \theta_{j+\half}}
             + \frac{B^{*\theta}_{i\ j-\half}}{\sin \theta_{j-\half}} 
       \right)
\end{eqnarray}
for the cell-centered magnetic-field components. This prescription
guarantees that for a homogeneous field parallel to the rotation axis,
the magnetic pressure is equal at the cell center and the cell
interface. It also increases the stability of the code for highly
magnetized flows, especially near MHD equilibria, and is critical
for the success of some of the tests presented here.

\subsection{Recovery of primitive variables}
In relativistic hydrodynamics, in contrast to the Newtonian case,
there exists no explicit expression
for the primitive variables ($\rho, v^i, \epsilon$) in terms of the
conserved ones ($D, S_i, \tau$). Hence,
a recovery procedure is required whereby the primitive variables are
obtained from the conserved ones by inverting the nonlinear system
given by Eqs.\,(\ref{eq:def:d}\,-\,\ref{eq:def:tau}) with an efficient
numerical algorithm. In most of the recovery algorithms
\citep{noble2006}, one first introduces some scalar quantities, and
then solves the resulting simplified system of equations before
recovering the primitives.

Following \cite{anton06}, our recovery procedure is based on the two
scalar quantities (note that the first of these is the conserved energy)
\begin{eqnarray}
\tau &=& \rho h W^2 - P + b^2 (W^2 - 1/2) - \alpha^2 (b^0)^2 - D,
\label{eq:tau}\\
S^2 &\equiv& \gamma^{ij} S_i S_j  = (\rho h + b^2)^2 W^4 v^2 \nonumber\\
&&+ \alpha^2 (b^0)^2
\left[ -2\rho h W^2 + b^2 (1 - 2 W^2) + \alpha^2 (b^0)^2 \right].
\label{eq:s2}
\end{eqnarray}
If one defines $z\equiv\rho h W^2$ and makes use of the expression
$\vec{B}\cdot\vec{S} = \rho h W \alpha b^0$ to eliminate $b^0$ from
these equations, the resulting expressions
\begin{eqnarray}
\left[ (z+B^2)^2 - S^2 - \frac{2z + B^2}{z^2} 
     (\vec{B}\cdot\vec{S})^2\right] W^2 
 &&\nonumber \\ - (z + B^2)^2 &=& 0, 
  \label{eq:rec_f1}\\
\left[ \tau + D - z - B^2 + \frac{(\vec{B}\cdot\vec{S})^2}{2z^2} 
       + P\right] W^2 + \frac{B^2}{2} &=& 0, \nonumber \\ 
  \label{eq:rec_f2}
 \end{eqnarray}
depend only on conserved quantities, on the metric, and on the set of
unknowns $\{P, z, W\}$, respectively. The system formed by
Eqs.\,(\ref{eq:rec_f1}\,-\,\ref{eq:rec_f2}) and the EOS can then be
solved to obtain $\{P, z, W\}$. From these three quantities, the
primitive variables can be easily computed as
\begin{eqnarray}
\rho     &=& \frac{D}{W} \label{eq:pzw2prim_1},\\
 v^i     &=& \frac{\gamma^{ij} S_j + (\vec{B}\cdot\vec{S}) B^i/ z}{z + B^2},
             \label{eq:pzw2prim_2}\\
\epsilon &=& \frac{z - D W - P W^2}{D W}. \label{eq:pzw2prim_3}
\end{eqnarray}
The numerical procedure to solve the system of
Eqs.\,(\ref{eq:rec_f1}\,-\,\ref{eq:rec_f2}) therefore depends on the EOS
(see next two subsections).

\subsubsection{Barotropic fluid}
In a barotropic fluid, the pressure depends only on the density, \ie
$P (\rho)$. In many astrophysical situations (\eg cold neutron
stars) as well as in many standard tests of numerical codes, the fluid
is assumed to be barotropic. The most commonly used barotropic EOS is
the polytropic EOS, $P = K \rho^\Gamma$, where $K$ is the polytropic
constant and $\Gamma$ is the adiabatic index.

For a barotropic EOS, the enthalpy is a function of the density only,
\ie $h(\rho)$, and thus $z = D h(\rho) W$. Using this fact and
Eq.\,(\ref{eq:pzw2prim_1}), it is possible to eliminate the unknowns
$P$ and $z$ from Eqs.\,(\ref{eq:rec_f1}\,-\,\ref{eq:rec_f2}). 
The Lorentz factor $W$ then remains to be computed numerically
by solving one of these equations.

Following \cite{antonphd}, we solve Eq.\,(\ref{eq:rec_f1}) for $W$ by
means of the bisection method, and then recover $P$ and $z$ using the
EOS. This method is extremely robust and always leads to a solution for $W$,
provided that it lies between the initial lower and upper guess value.

\subsubsection{Baroclinic fluid}
This is the most common form of the EOS in hydrodynamic simulations
because it takes into account temperature effects. We implemented
several baroclinic EOS in our numerical code, namely the ideal gas
EOS, $P=\rho\epsilon(\Gamma-1)$, the analytic hybrid EOS
\citep{janka_93_a}, and a tabulated microphysical EOS.  

Irrespective of the baroclinic EOS used, it can always be expressed 
in the form $P (\rho, \epsilon, Y_i)$. Since the composition $Y_i$
(the index $i$ runs over all relevant species) is known directly from
the hydrodynamics, the dependence of the EOS on the composition does
not affect the recovery procedure. The following discussion therefore can
be restricted to an EOS of the form $P (\rho, \epsilon)$.

For a baroclinic fluid, the system formed by
Eqs.\,(\ref{eq:rec_f1}\,-\,\ref{eq:rec_f2}) and the EOS expressed as
\begin{equation}
P - P (\rho, \epsilon) =0 \label{eq:eos_nr}
\end{equation}
must be solved numerically. In general, it is not possible to use
the EOS to analytically remove the dependence of
Eqs.\,(\ref{eq:rec_f1}\,-\,\ref{eq:rec_f2}) on $P$, since $\epsilon$
depends on the pressure itself due to Eq.\,(\ref{eq:pzw2prim_3}).
However, for some analytic EOS, \eg an ideal gas,
Eqs.\,(\ref{eq:pzw2prim_1}) and (\ref{eq:pzw2prim_3}) can be used to
express the pressure as a function of $z$ and $W$ only. This allows
one to eliminate $P$ from Eqs.\,(\ref{eq:rec_f1}\,-\,\ref{eq:rec_f2}),
reducing the system to be solved to two equations with the unknowns
$z$ and $W$. Since the numerical method should not rely on any
assumption about the EOS, we do not exploit this
simplification. Instead we solve the system of Eqs.\,(\ref{eq:rec_f1},
\ref{eq:rec_f2}, \ref{eq:eos_nr}) by means of a Newton-Raphson scheme,
which converges rapidly provided the initial guess is sufficiently good (see
below). We consider the Newton-Raphson iteration to be converged when
the relative error of the variables is less than a certain tolerance
(typically $10^{-12}$).

\paragraph{Microphysical tabulated EOS.}
Some of the equations of state available from nuclear physics that are used
in astrophysics are not provided in terms of the specific internal
energy. In general, the EOS depends on the composition of the fluid
(usually the electron fraction $Y_e$) and on the temperature $T$
instead of $\epsilon$. Hence, one has to deal with tabulated EOS of
the form
\begin{eqnarray}
P &=& P (\rho, T, Y_e) ,\\
\epsilon &=& \epsilon (\rho, T, Y_e).
\label{eq:eps_rty}
\end{eqnarray}
A first approach to handle such an EOS is to obtain effectively 
$P=P(\rho,\epsilon,Y_e)$ by computing the value of $T$ that satisfies
Eq.~(\ref{eq:eps_rty}) for a given value of $\epsilon$. The
procedure described above for an EOS of the form $P (\rho, \epsilon)$
can then be applied, since $Y_e$ is known directly from the evolution.
This approach was successfully used in the hydrodynamic simulations
presented by \cite{ott07b}. In the magnetized case, however, we find
this approach to be problematic for strong magnetic fields ($P_{\rm
mag}/P > 1$). The solver is able to recover the exact value in that
regime only if the initial guess is very close to the solution, which
renders the code unstable.  To avoid this problem we add the equation 
\begin{equation}
\epsilon - \epsilon (\rho, T, Y_e) = 0,
\label{eq:eps_nr}
\end{equation}
to the Newton-Raphson system, and solve the extended system of 
Eqs.~(\ref{eq:rec_f1}, \ref{eq:rec_f2}, \ref{eq:eps_nr}) for the unknowns
$z$, $W$, and $T$.  This allows one to use directly the EOS as a
function of $T$ instead of $\epsilon$. We find that this method is
very stable and has a much larger radius of convergence than the
first approach (see Sect.~\ref{sec:recovertest}).

\paragraph{``Safe'' guess values.}
\label{sec:safeguess}
When we use the values of the previous time step
as an initial guess for the Newton-Raphson iteration at a
given time step, the solver
usually converges within a few iterations. However, sometimes the
guess values are too far away from the solution, and the
Newton-Raphson iteration fails. In such a case, we restart the
iteration process using a ``safe'' set of guess values, which we
choose to be upper limits to the unknowns $\{P,z,W\}$ (or $\{T, z,
W\}$). This choice leads to a rather robust recovery scheme as
demonstrated by our test calculations (see Sect.\,\ref{sec:tests}).

To derive an upper limit for $z$, we define $\delta$ as the
angle between $\vec{v}$ and $\vec{B}$, \ie $\vec{v}\cdot\vec{B} =
\sqrt{\vec{v}^2\vec{B}^2} \cos \delta$. Using this angle, we have
\begin{eqnarray}
&& \alpha^2 (b^0)^2 = W^2 \vec{B}^2\vec{v}^2 \cos^2 \delta, \\
&& b^2 = \frac{\vec{B}^2}{W^2}\left( 1 + (W^2-1) \cos^2 \delta \right).
\end{eqnarray}
From the definition of $\tau$ given in Eq.\,(\ref{eq:def:tau}), one
obtains $z$ as
\begin{equation}
z = \tau + P + D - \frac{B^2}{2} - \frac{B^2}{2} 
    \frac{W^2-1}{W^2} (2-\cos^2 \delta).
\label{eq:app:z}
\end{equation}
Since the last term in this equation is always negative or zero, an
upper limit for $z$ is given by
\begin{equation}
z \le \tau + P + D - \frac{B^2}{2}.
\end{equation}
However, this upper limit cannot be computed directly from the
conserved quantities, since the pressure is unknown. Hence, we first
need to determine an upper limit for the pressure.  If we assume that
the pressure grows monotonically with $\rho$ and $\epsilon$, which is
a reasonable assumption for the types of EOS that we use in the code, then
we only need to derive upper limits for $\rho$ and $\epsilon$, and
hence
\begin{equation}
P \le P_{\rm max} \equiv P (\rho_{\rm max}, \epsilon_{\rm max}).
\end{equation}
It is easy to find an upper limit for $\rho$, since $W\ge 1$,
\begin{equation}
\rho \le \rho_{\rm max} \equiv D.
\end{equation}
In the case of the specific internal energy, we substitute
Eq.\,(\ref{eq:app:z}) into Eq.\,(\ref{eq:pzw2prim_3}) and obtain
\begin{eqnarray}
\epsilon &= \frac{1}{DW} &\left[ \tau  - \frac{B^2}{2} 
            + P(1-W^2) + D (1-W) \right. \nonumber\\
&&\left.- \frac{B^2}{2} \frac{W^2-1}{W^2} (2-\cos^2 \delta) \right].
\end{eqnarray}
Using again the fact that $W\ge 1$ we derive the upper limit
\begin{equation}
\epsilon \le \epsilon_{\rm max} \equiv 
\frac{1}{D} \left[ \tau  - \frac{B^2}{2}  \right].
\end{equation}
An upper limit for $z$ is given by
\begin{equation}
z \le z_{\rm max} \equiv \tau + P_{\rm max} + D - \frac{B^2}{2},
\end{equation}
which coincides with $z$ in the limit of small velocities, $W\to 1$.

We were unable to compute an analytic upper limit for $W$; however
it is easy to set an upper limit from physical considerations. In the
core-collapse simulations in which we are interested, the Lorentz factor is
not expected to exceed a value of $10$. Nevertheless, we chose a
much larger guess value for $W$, since the number of iterations until
convergence is very insensitive to the precise value of the upper
limit.  Accordingly, the ``safe'' guess values that we use in the
Newton-Raphson solver are
\begin{eqnarray}
P_{\rm guess} &=& P_{\rm max}, \\
z_{\rm guess} &=& z_{\rm max}, \\
W_{\rm guess} &=& 10000, \\
T_{\rm guess} &=& T (\rho_{\rm max}, \epsilon_{\rm max}).
\end{eqnarray}
We note that $T_{\rm guess}$ is not an upper limit for the temperature in
general, but the pressure value computed with $T_{\rm guess}$ and
$\rho_{\rm max}$ provides an upper limit for the pressure, \ie $P
\le P (\rho_{\rm max}, T_{\rm guess})$.

\subsection{Vacuum treatment}
\label{sec:vacuum}
The presence of vacuum regions is common in numerical
simulations dealing with astrophysical scenarios. These regions are
usually avoided by imposing a numerical atmosphere surrounding the
object under study, \ie a small floor value for the rest mass
density, which allows one to use the same recovery procedure in
regions filled with the numerical atmosphere and the fluid. A vacuum
region would cause the recovery procedure to fail both in the
hydrodynamic and in the magneto-hydrodynamic case, as can be inferred
from Eqs.\,(\ref{eq:pzw2prim_2},\,\ref{eq:pzw2prim_3}). The
numerical atmosphere approach is commonly used in hydrodynamic
simulations \citep[see \eg][]{font_02_a, dimmelmeier_02_a} as well as
in GRMHD simulations \citep{Duez05, shibata05, giacomazzo07}.

In the unmagnetized case, the floor value of the numerical atmosphere
is chosen such that it does not affect significantly the dynamics of
the system. This can be achieved by choosing the threshold value for
the rest mass density to be a small fraction of the maximum density in
the initial model, typically $\rho_{\rm thr} \approx 10^{-6} \rho_{\rm
max}$.  Every grid point with $\rho < \rho_{\rm thr}$ is reset to the
numerical atmosphere value, \ie $\rho = \rho_{\rm atm}$ and $v^i=0$,
where the floor value for the rest mass density is $\rho_{\rm atm}
\approx 10^{-3} \rho_{\rm thr}$.

In the magnetized case additional problems arise. Since the transition
to the numerical atmosphere usually results in a steep profile in $\rho$
(which drops to the floor value) but not necessary in $B$ (magnetic
field lines can extend into the vacuum), atmosphere regions can easily
have large ratios of $P_{\rm mag}/P$, even if the fluid is weakly
magnetized. This problem increases as the floor value $\rho_{\rm atm}$
is reduced, and can lead to problems with the recovery of the
primitive variables in the atmosphere.  To avoid this problem, some
authors \citep{Duez05,shibata05} do not allow magnetic fields in the
numerical atmosphere by choosing magnetic fields confined to the fluid
regions. Other authors \citep{giacomazzo07, shibata07} apply a floor to
the hydrodynamic variables and allow the magnetic field to evolve freely.
This approach works fine, if the ratio of $P_{\rm mag}/P$ does not
exceed the critical value above which the recovery procedure fails. We
estimate this critical value for our code in
Sect.\,\ref{sec:recovertest}.  Consequently, a sufficiently low
density atmosphere will show the correct dynamic behavior when the
magnetic field strength in the atmosphere is limited.  On the
other hand, if one wishes to simulate stronger magnetic fields, one must
use a denser atmosphere that can even affect the dynamics of the
system \citep{shibata07}.

We also allow for a freely evolving magnetic field in the atmosphere,
since we are then not restricted to any particular magnetic field structure.  
To overcome the problem of the high magnetization $P_{\rm mag}/P$ in the
atmosphere, we choose compromise values for $\rho_{\rm atm}$ depending
on the problem to be solved.  Since there are cases (\eg for the
evolution of neutron stars in Sect.\,\ref{sec:mns}, $P_{\rm mag}/P$
becomes as large as $10^{13}$) where the ratio $P_{\rm mag}/P$ exceeds
the critical value for the recovery procedure ($10^6-10^8$; see
Sect.\,\ref{sec:recovertest}) even for reasonable values of $\rho_{\rm
atm}$, we use a fast atmosphere checking routine which avoids the
recovery of the primitives in the respective zones.

If we are able to mark a zone as being part of the atmosphere before
the recovery of the primitives is performed, we can avoid the recovery
because the values for the primitives in these zones are set to the
floor value.  Since $D \ge \rho$, if $D < \rho_{\rm thr}$ holds, then
$\rho < \rho_{\rm thr}$, and the zone is part of the atmosphere.
Hence, we can use this condition to check whether a zone belongs to
the atmosphere before performing the recovery.  We note that for
atmosphere zones, whose velocities are set to zero at every time step,
it is very unlikely that the value of the (unknown) Lorentz factor $W$
at the next time step differs significantly from $1$, \ie $\rho \approx
D$ in these zones. Using this procedure, we can handle arbitrarily
large magnetic fields in the atmosphere without imposing any
limitation on the value of $\rho_{\rm atm}$.

In the equilibrium models of magnetized neutron stars
(Sect.\,\ref{sec:mns}), we keep the magnetic field fixed in the
atmosphere to its initial value. This is a reasonable choice since for
an equilibrium configuration the outside magnetic field should not
change during the evolution. The advantage of this approach is that
the time step is not dominated by the atmosphere where the eigenvalues
are close to the speed of light, but by the neutron star interior with
eigenvalues of the order of $\approx 0.1$. This results in a speed-up
of about a factor of $10$ in these computations. We note that this speed-up 
is only possible because in the CFC approximation the metric
evolution does not constrain the size of the time step. In codes based
on hyperbolic formulations of the Einstein equations, the time step is
always limited by the light-crossing time of the zones, \ie this
speed-up is impossible. All other existing GRMHD codes with dynamic
spacetimes \citep{Duez05,shibata05,giacomazzo07} suffer from this
limitation.

\begin{table}[t!]
  \centering
  \caption{Test of the recovery of the primitive variables. Varying
           the flow velocity $\vec{v}$, the magnetic field $\vec{B}$,
           and the angle between $\vec{B}$ and $\vec{v}$ in a wide
           range, the recovery procedure is tested for the equations
           of state, the densities $\rho$, the specific internal
           energies $\epsilon$, and the electron fractions $Y_{\rm e}$
           given in columns 2 to 5, respectively.}
  \begin{tabular}{llccc}
    \hline \hline
    Test & EOS & $\rho [g\,cm^{-3}]$ & $\epsilon$ & $Y_{\rm e}$
    \\
    \hline
    PN  & Polytropic & $3 \times 10^{14}$  & -       & - \\
    PE  & Polytropic & $10^{12}$           & -       & - \\
    HN1 & Hybrid     & $3 \times 10^{14}$  & 0.001   & - \\
    HN2 & Hybrid     & $3 \times 10^{14}$  & 0.01    & - \\
    HN3 & Hybrid     & $3 \times 10^{14}$  & 0.1     & - \\
    HN4 & Hybrid     & $3 \times 10^{14}$  & 1.0     & - \\
    HE1 & Hybrid     & $10^{12}$           & 0.001   & - \\
    HE2 & Hybrid     & $10^{12}$           & 0.01    & - \\
    HE3 & Hybrid     & $10^{12}$           & 0.1     & - \\
    HE4 & Hybrid     & $10^{12}$           & 1.0     & - \\
    I1  & Ideal gas  & $10^{14}$           & 0.01    & - \\
    I2  & Ideal gas  & $10^{14}$           & 0.1     & - \\
    I3  & Ideal gas  & $10^{14}$           & 1.0     & - \\
    I4  & Ideal gas  & $10^{14}$           & 10.0    & - \\
    I5  & Ideal gas  & $10^{14}$           & 100.0   & - \\
    I6  & Ideal gas  & $10^{14}$           & 1000.0  & - \\
    I7  & Ideal gas  & $10^{14}$           & 10000.0 & - \\
    S1  & SHEN       & $2.4\times 10^{14}$ & 0.055   & 0.25  \\
    S2  & SHEN       & $4.2\times 10^{9} $ & 0.009   & 0.427 \\
    S3  & SHEN       & $4.2\times 10^{8} $ & 0.008   & 0.457 \\
    S4  & SHEN       & $2.6\times 10^{6} $ & 0.009   & 0.5   \\
    \hline \hline
  \end{tabular}
  \label{tab:recovery}
\end{table}

\section{Code tests}
\label{sec:tests}
%
\subsection{Recovery of the primitive variables}
\label{sec:recovertest}
The numerical method used for the recovery of the primitive variables
$\{\rho, v^i, \epsilon, B^j\}$ from the conserved ones $\{D, S_i,
\tau, B^j\}$, is tested by varying the flow velocity $\vec{v}$, the
magnetic field $\vec{B}$, and the angle $\delta$ between $\vec{B}$ and
$\vec{v}$ over a wide range. Instead of varying $\vec{v}$ and
$\vec{B}$ we vary the Lorentz factor $W-1$ in the interval
$[10^{-4}, 10^4]$, and the magnetization $P_{\rm mag} / P$ in the
interval $[10^{-8}, 10^{10}]$, respectively.  We choose values of
$\rho$, $\epsilon$, and (for the tabulated EOS only) $Y_{\rm e}$ that
are typical for core collapse (Table~\ref{tab:recovery}). Test cases PN
and PE correspond to a polytropic EOS with $\Gamma = 2$, $K =
1.455\times 10^5$ (cgs units), and $\Gamma = 4/3$, $K = 4.897\times
10^{14}$, respectively. For the hybrid EOS, the test cases HN and HE
probe densities above and below nuclear matter density, while the SHEN
EOS cases test the typical conditions inside a proto-neutron star (S1)
and the progenitor core (S2 to S4).

\begin{figure}[t!]
  \centering
  \resizebox{0.47\textwidth}{!}{\includegraphics*{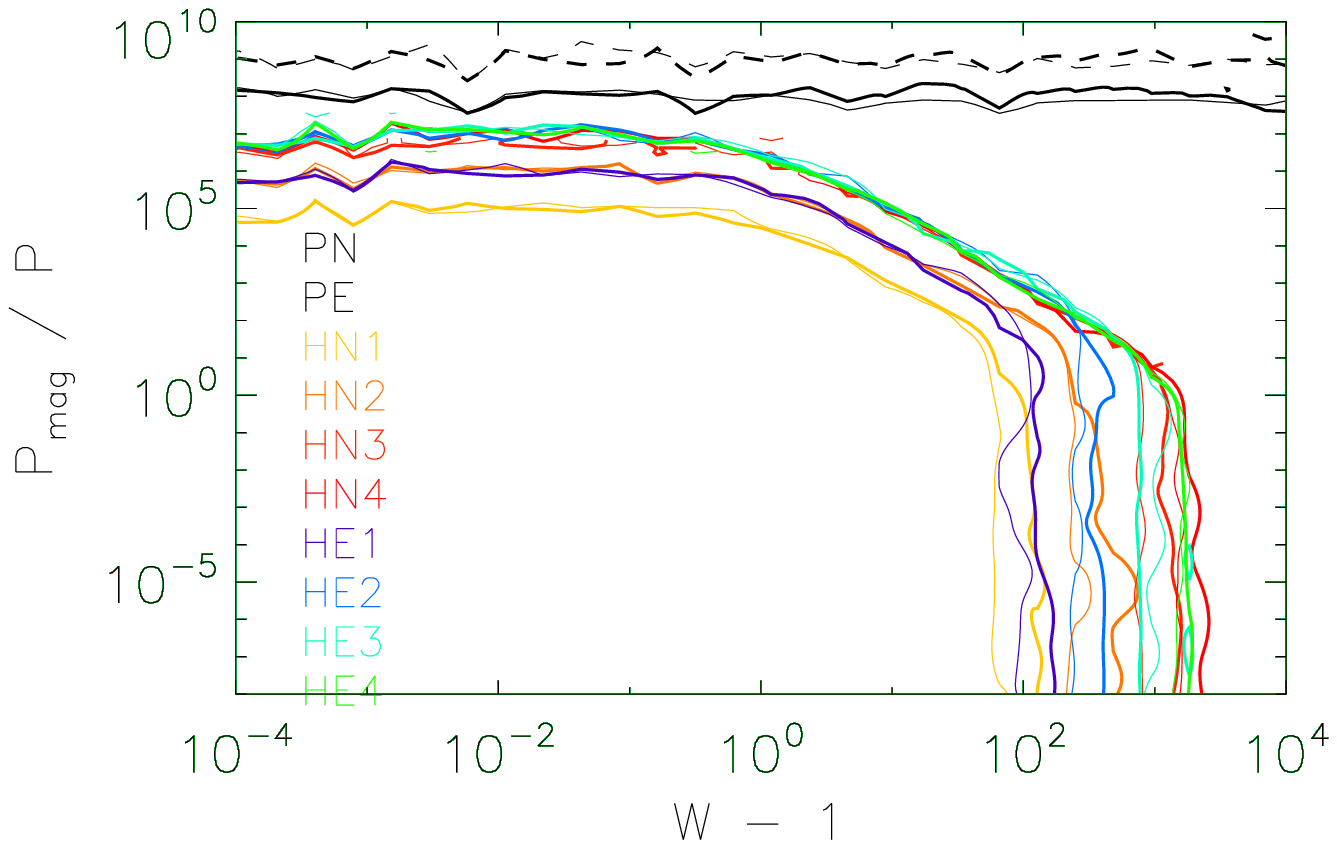}}
  \\ [0.5 em]
  \resizebox{0.47\textwidth}{!}{\includegraphics*{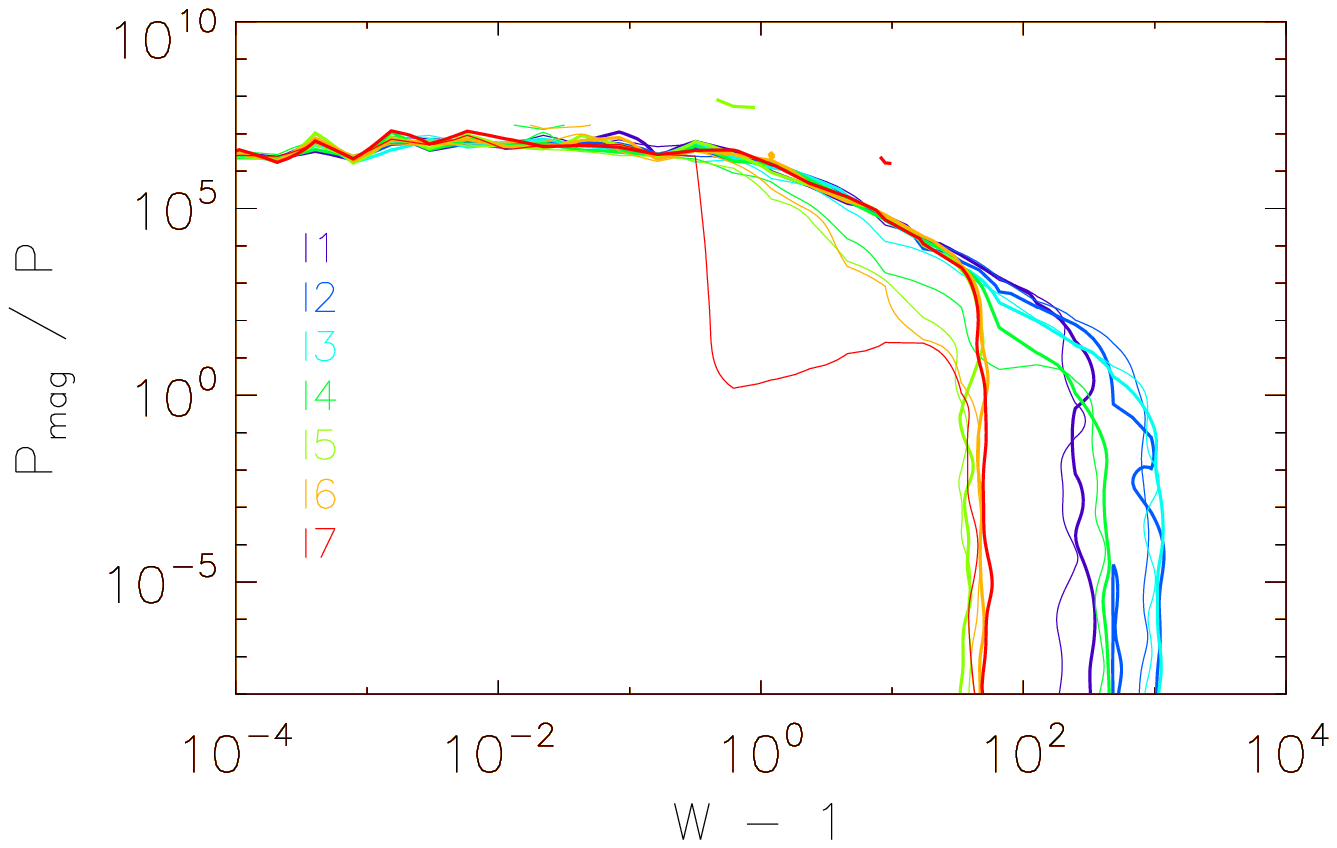}}
  \\ [0.5 em]
  \resizebox{0.47\textwidth}{!}{\includegraphics*{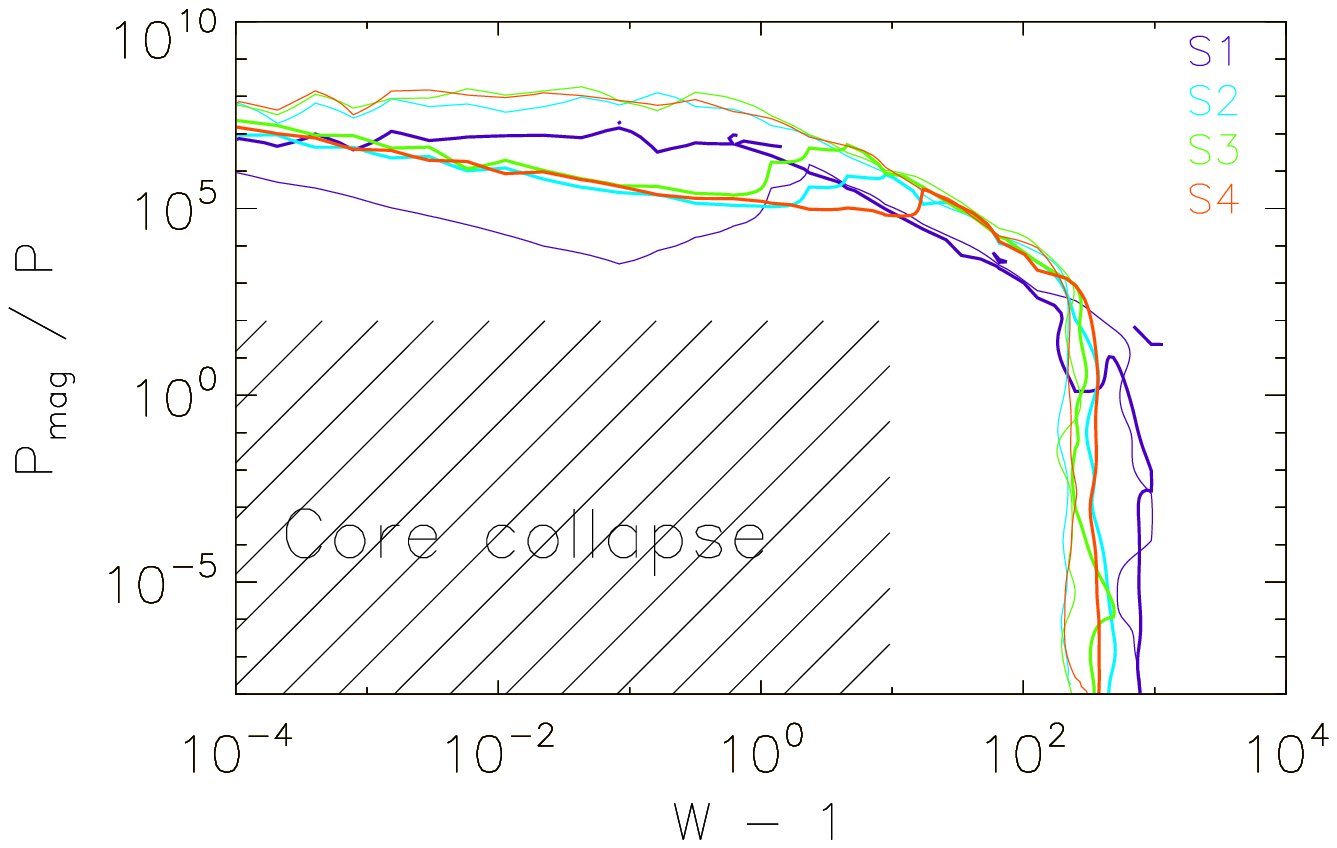}}
  \caption{Upper limits for the Lorentz factor ($W-1$) and the
           magnetization $P_{\rm mag}/P$ for which the relative
           difference between the values of the recovered primitive
           variables and their exact values is less than
           $10^{-10}$. The upper panel shows these limits for the
           polytropic and the hybrid EOS, the middle panel for the
           ideal gas EOS, and the lower panel for the tabulated SHEN
           EOS, respectively. Thin lines correspond to the case
           $\cos^2 \delta = 0$, and thick lines to $\cos^2 \delta =
           1.0$. The shaded region in the bottom panel shows the
           typical parameter space encountered in core-collapse
           simulations.}
  \label{fig:recovery}
\end{figure}

For a baroclinic EOS, we use the ``safe'' values given in
Sect.\,\ref{sec:safeguess} as guess values for the Newton-Raphson
iteration, and choose a tolerance of $10^{-12}$.
Figure~\ref{fig:recovery} shows the region in which the recovery scheme
converges well, \ie where the relative difference between the values
of the recovered primitive variables and their exact values is less
than $10^{-10}$. For all considered EOS, the parameter space
of astrophysical interest is well covered.  In test case I7, the region
of convergence is reduced substantially when $\cos \delta = 0$. This
test, that corresponds to an extreme ideal gas with $\epsilon =
10^{4}$, was chosen to determine the maximum value of $\epsilon$ which
can be recovered and lies outside the parameter range of interest in
core collapse.
We also determine upper
limits to the magnetization $P_{\rm mag}/P$ in the low velocity limit
ranging from values of $10^6$ to $10^8$. 
In the low magnetic field limit the maximum Lorentz factor
that the recovery procedure can handle is $10^2$ to $10^3$. 
If any of these limits is exceeded, the numerical
scheme is unable to recover the primitive quantities within the
required accuracy.  
The reason for the recovery failure in the three limiting cases for
$\epsilon$, $W$, and $P_{\rm mag}/P$ is that the contribution of
internal energy, kinetic energy or magnetic energy, respectively, is
dominant in the system and any other kind of energy has a very
small contribution. In these cases large changes in the subdominant terms
will produce small changes in the recovery equations and
the system may therefore converge to a wrong solution within a given accuracy.
If the tolerance value is reduced, these limits can
be extended. We note that in astrophysical situations involving
baryonic matter, it is unlikely to encounter values of $\epsilon > 1$,
$W > 10$, or $P_{\rm mag}/P > 100$. Therefore, we consider that our
recovery procedure is sufficiently robust for simulations
of core collapse and involving compact objects.

\begin{table*}[t!]
  \centering 
  \caption{Spherical explosion test. Initial values for the pressure
           $P$, the density $\rho$, the magnetic field $|\vec{B}|$,
           and the magnetization $P_{\rm mag}/P$ are given in columns
           2 to 5 for the explosion region ($r<1$) and in the ambient
           region ($r>1$), respectively. The eigenvalues, namely the
           speeds of the fast magnetosonic wave $\lambda_{f\pm}$, the
           Alfv\'en wave $\lambda_{A\pm}$, the slow magnetosonic wave
           $\lambda_{s\pm}$, and of the entropy wave $\lambda_e$ in
           the radial direction at the initial time are given in
           columns 6 to 9, both for the equator (left value) and the
           pole (right value).  In addition, we provide in the last
           column the value of the eigenvalue (sound speed)
           $\lambda_{\pm}$ of the corresponding non-magnetized case.}
  \begin{tabular}{llllllllll}
    \hline \hline
          & $P$ & $\rho$ & $|\vec{B}|$ & $P_{\rm mag}/P$ 
          & \quad $|\lambda_{f\pm}|$ & \; $|\lambda_{A\pm}|$       
          & \; $|\lambda_{s\pm}|$ & $|\lambda_e|$ & $|\lambda_{\pm}|$\\
    \hline\\ [-0.7 em]
    $r<1$ & $1$ & $10^{-2}$ & $0.1$ & $5\times10^{-3}$
          & $0.578,\, 0.576$  & $0,\, 0.05$ 
          & $0,\, 0.05$ & $0$ & $0.576$\\
    $r>1$ & $3\times 10^{-5}$ & $10^{-4}$ & $0.1$  & $166.6$
          & $0.991,\, 0.989$ & $0,\, 0.988$ & $0,\, 0.426$ & $0$& $0.426$\\
    \hline \hline
  \end{tabular}
  \label{tab:SEparam}
\end{table*}

\begin{figure*}[t!]
  \centering
  \resizebox{0.48\textwidth}{!}{\includegraphics*{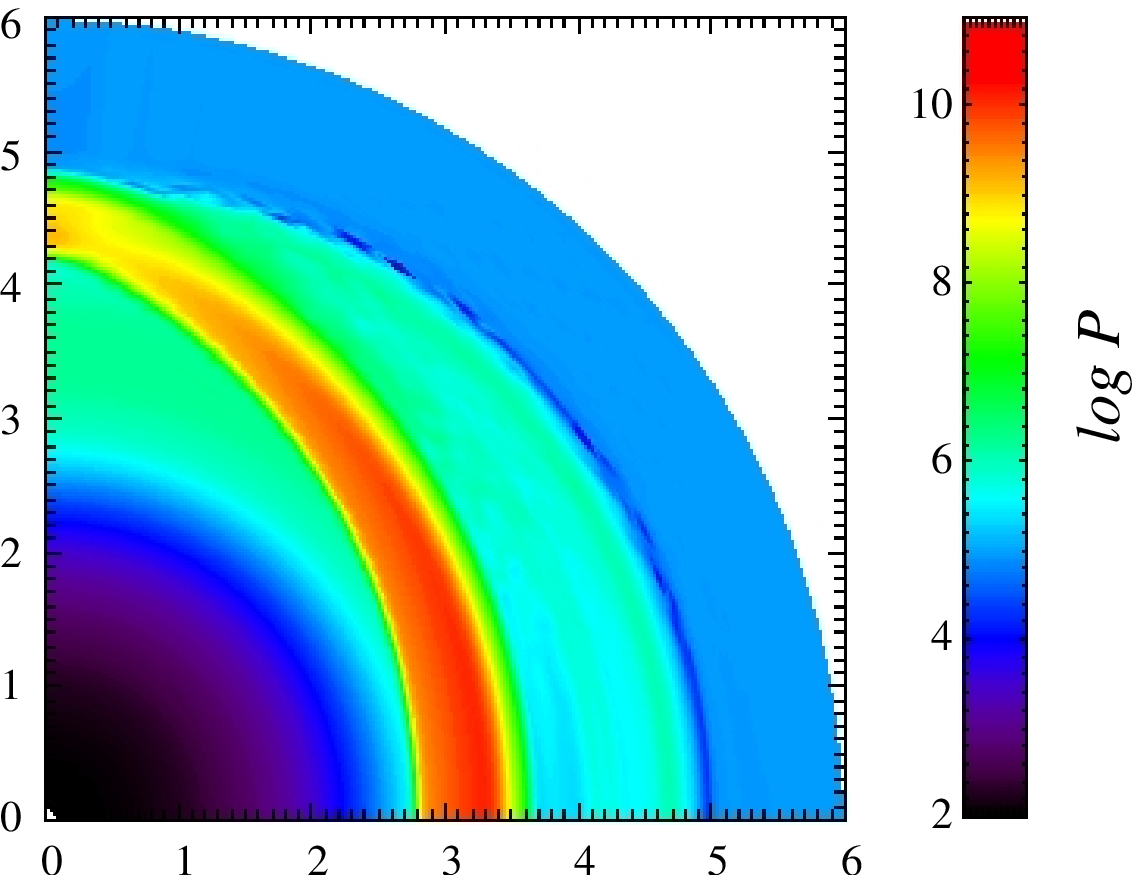}} \quad
  \resizebox{0.48\textwidth}{!}{\includegraphics*{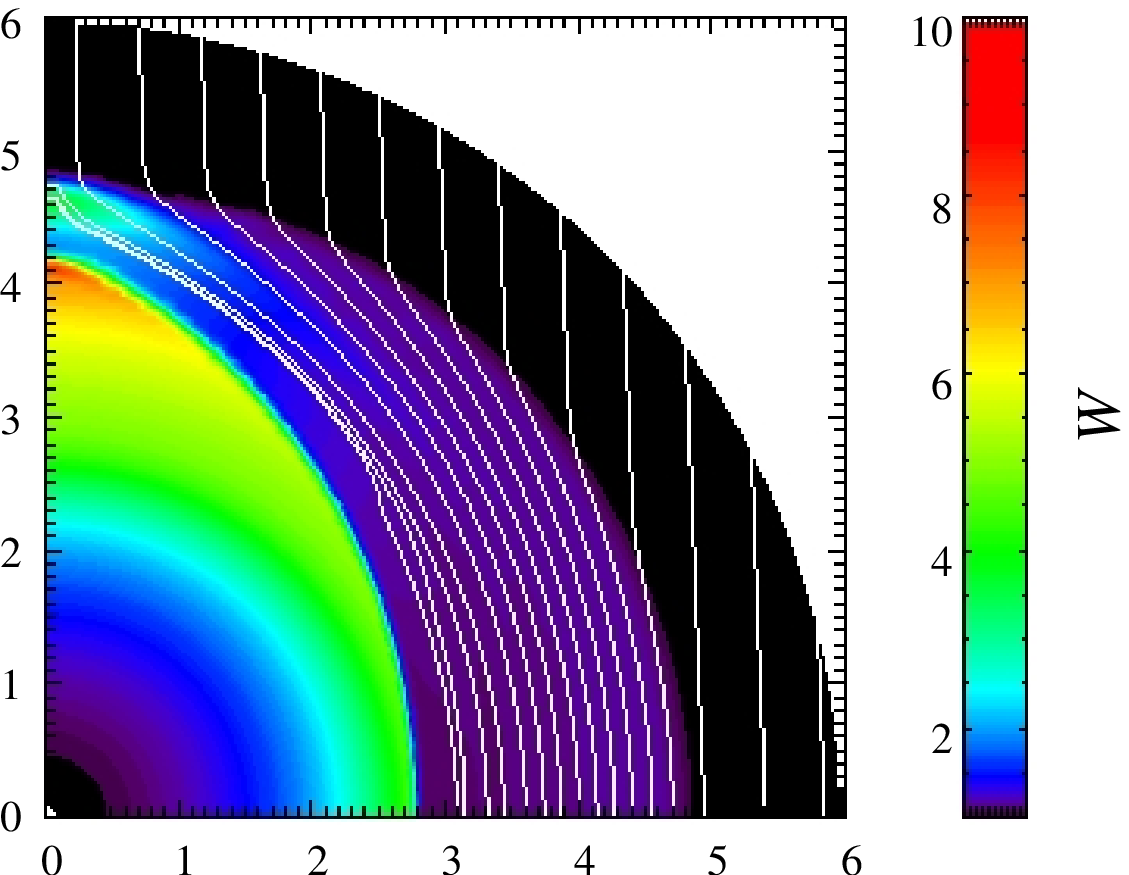}}
  \caption{Snapshot of the spherical explosion test at $t=4$. The
           panels show the logarithm of the pressure (left) and
           Lorentz factor, and the magnetic field lines
           (right). The simulation was performed with $160\times 40$
           zones using PHM cell reconstruction and the KT flux
           formula.}
  \label{fig:SEcolor}
\end{figure*}

Using the ``safe'' guess values, the number of iterations needed for
the Newton-Raphson solver to converge is relatively large: $50-70$ for
both the hybrid and the ideal gas EOS, and $50-200$ for the tabulated
EOS. However, during a numerical simulation, the ``safe'' guess is only
used if the regular guess (the value from the previous time step)
fails. If we use guess values that differ by only 10\% from the exact
ones, the Newton-Raphson converges more rapidly, within $10-20$ iterations 
for the hybrid and ideal gas EOS, and $20-30$ for the tabulated EOS. For the
polytropic EOS it takes about $40-60$ bisection steps to achieve the
required tolerance.

\subsection{Spherical explosion}
\label{sec:SE}
Since the majority of existing (2D) MHD codes are written in
cylindrical coordinates, a commonly performed test is the simulation
of a cylindrical explosion. For relativistic MHD codes, such a
setup was proposed by \citet{komissarov99}, which was also used by
other authors \citep{delzanna03, leismann05}. However, when using a
code based on spherical coordinates the most natural choice is a
spherical explosion.  \citet{koessl90} performed this test with a
Newtonian MHD code, but to the best of our knowledge no spherical explosions
test has been performed in relativistic MHD. Therefore, we consider
here a spherical explosion test for which the initial jump conditions are
identical to those of the cylindrical test of \citet{komissarov99}.

\begin{figure*}[t!]
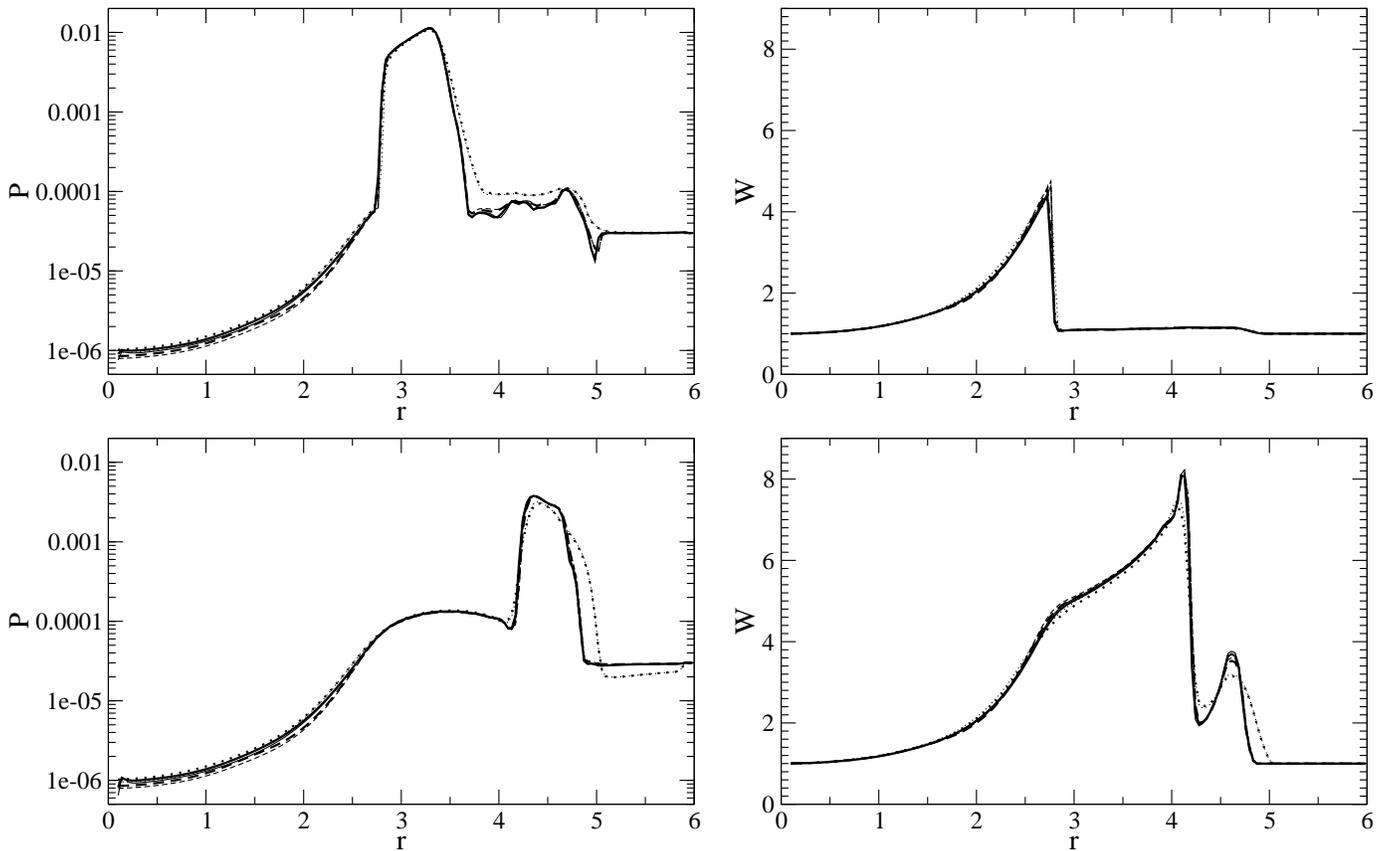

  \centering
  \resizebox{!}{0.3\textwidth}{\includegraphics*{0086f3a.eps}} \quad
  \resizebox{!}{0.3\textwidth}{\includegraphics*{0086f3b.eps}}
  \\ [0.5 em]
  \resizebox{!}{0.3\textwidth}{\includegraphics*{0086f3c.eps}} \quad
  \resizebox{!}{0.3\textwidth}{\includegraphics*{0086f3d.eps}}
  \caption{Results for the spherical test explosion at $t=4$. The
           panels show radial profiles of the fluid pressure $P$ (left
           panels) and the Lorentz factor $W$ (right panels) along the
           equator (upper panels) and the polar axis (lower panels),
           respectively.  The lines styles differentiate between the
           reconstruction schemes: minmod (dotted), MC (dashed), and
           PHM (solid). Results for two different flux formulae are
           shown: HLLE (thin lines) and KT (thick lines). Note that
           the results obtained with these two flux formulae are often
           so similar that they cannot be distinguished.  The grid
           resolution used is $160\times 40$ zones.}
  \label{fig:SEprofiles}
\end{figure*}

Our test setup consists of an initial radially symmetric explosion
zone ($r<1$) surrounded by a highly magnetized ambient gas for
$r>1$. In the outer part of the explosion region ($0.8 < r < 1.0$), we
set the state variables decline exponentially to the values of the
ambient medium (Table \ref{tab:SEparam}). The velocity is initially
zero everywhere, and the magnetic field is homogeneous and parallel to
the symmetry axis. The background spacetime is assumed to be flat. The
initial data are evolved using an ideal gas EOS with an adiabatic
index $\Gamma = 4 / 3$. The computational grid is evenly spaced in
radius and angle, and extends in the radial direction up to a maximum
radius of $r = 6.0$. We perform the test with two $(r,\theta)$
resolutions ($ 80 \times 20 $, and $ 160 \times 40 $) for all
reconstruction schemes and flux formulae.

Table \ref{tab:SEparam} shows the eigenvalue structure of the initial
setup.  Since the explosion region is weakly magnetized ($P_{\rm
mag}/P = 5\times 10^{-3}$), the dominant wave in this region is the
fast magnetosonic wave, which propagates at a speed $|\lambda_{f}|$
close to that of the corresponding hydrodynamic wave. All other
wave speeds are close to zero, \ie the inner region will expand with
the velocity $|\lambda_{f}|$. The ambient gas is highly magnetized,
and the full wave structure is significant there. The only waves fast
enough to travel ahead of the explosion shock are the fast
magnetosonic wave, which propagates with an almost angular-independent
radial velocity close to the speed of light, and the Alfv\'en wave,
whose radial velocity is also close to $c$ along the symmetry axis,
but varies as $\cos \theta$.

Figure~\ref{fig:SEcolor} shows a snapshot at the end of the simulation
($t=4$). Both the Lorentz factor and the pressure distribution
clearly show the wave structure mentioned above. The spherical fast
magnetosonic wave is located at $r \approx 5$, and the trailing strong
shock, which is deformed due to the magnetic field, consists of a
mixture of the bulk expansion of the inner region and an Alfv\'en wave
propagating faster along the axis. Even further inwards, a rarefaction
wave is visible, which is almost spherically symmetric since the
magnetization in this region is rather low.

The corresponding radial profiles of $P$ and $W$ both along the
equator (upper panels) and the axis (lower panels) are displayed in
Fig.\,\ref{fig:SEprofiles} for various numerical methods. These plots
are qualitatively similar to those of the cylindrical explosion test
\citep[see \eg Fig.\,B.4 in][]{leismann05}. All numerical schemes
exhibit first order convergence as expected for flows with shocks. The
MC and PHM schemes yield very similar results, while the minmod scheme
gives slightly smaller values. No significant differences are found
between the results obtained using the HLLE and KT flux formulae.

\subsection{Magnetized neutron stars}
\label{sec:mns}
The previous two tests demonstrate the ability of the code to
handle extreme situations such as high magnetization, large Lorentz
factors, and strong shocks. In this section we show its correct behavior
in curved spacetimes, particularly in dynamic ones. An astrophysical
scenario that can be used for this assessment is the evolution of
equilibrium neutron stars, a test which is frequently used for general
relativistic hydrodynamics codes \citep{shibata_99, font_02_a,
dimmelmeier_02_a, duez_03, cerda05} as well as for GRMHD codes
\citep{giacomazzo07}. Of all presently existing codes capable of solving
the GRMHD equations coupled to a dynamic spacetime \citep{Duez05,
shibata05, giacomazzo07, anderson08}, this demanding test, involving
all aspects of the code and in particular the correct coupling between
metric and MHD equations, has only been performed by the 
code of  \cite{giacomazzo07}.

\begin{table*}[t!]
  \centering 
  \caption{Initial models of magnetized neutron stars. From left to
           right, the columns give the central current density $j_0$,
           the equatorial radius $r_{\rm e}$, the ratio of polar to
           equatorial radius $r_{\rm p}/r_{\rm e}$, the ratio of
           magnetic to thermal pressure $P_{\rm mag} / P$ at the
           center of the star, the central magnetic field
           $|\vec{B}|_{\rm c}$, and the ADM mass $M_{\rm ADM}$ of each
           model, respectively.}
  \begin{tabular}{lccccccc}
    \hline \hline \\ [-.9 em]
    Model & $j_0$ [A m$^{-2}$]& $r_{\rm e}$ [km] & $r_{\rm p}/r_{\rm e}$ & $P_{\rm mag} / P|_{\rm c}$ 
    & $|\vec{B}|_{\rm c}$ [$\sqrt{4\pi}$~G]& $M_{\rm ADM}$ [$M_{\odot}$] \\
    \hline \\ [-.9 em]
    MNS0 & 0                 & 11.998 & 1.0      &  & 0 & 1.40 \\
    MNS1 & $2\times 10^{13}$ & 11.998 & 0.999992 & $5.75\times 10^{-6}$ & $7.2\times 10^{14}$ & 1.40 \\
    MNS2 & $2\times 10^{14}$ & 11.999 & 0.9992   & $5.76\times 10^{-4}$ & $7.2\times 10^{15}$ & 1.40 \\
    MNS3 & $5\times 10^{14}$ & 12.006 & 0.995    & $3.63\times 10^{-3}$ & $1.8\times 10^{16}$ & 1.40 \\
    \hline \hline
  \end{tabular}
  \label{tab:MNS}
\end{table*}

\begin{figure}[t!]
  \centering
  \resizebox{0.40\textwidth}{!}{\includegraphics*{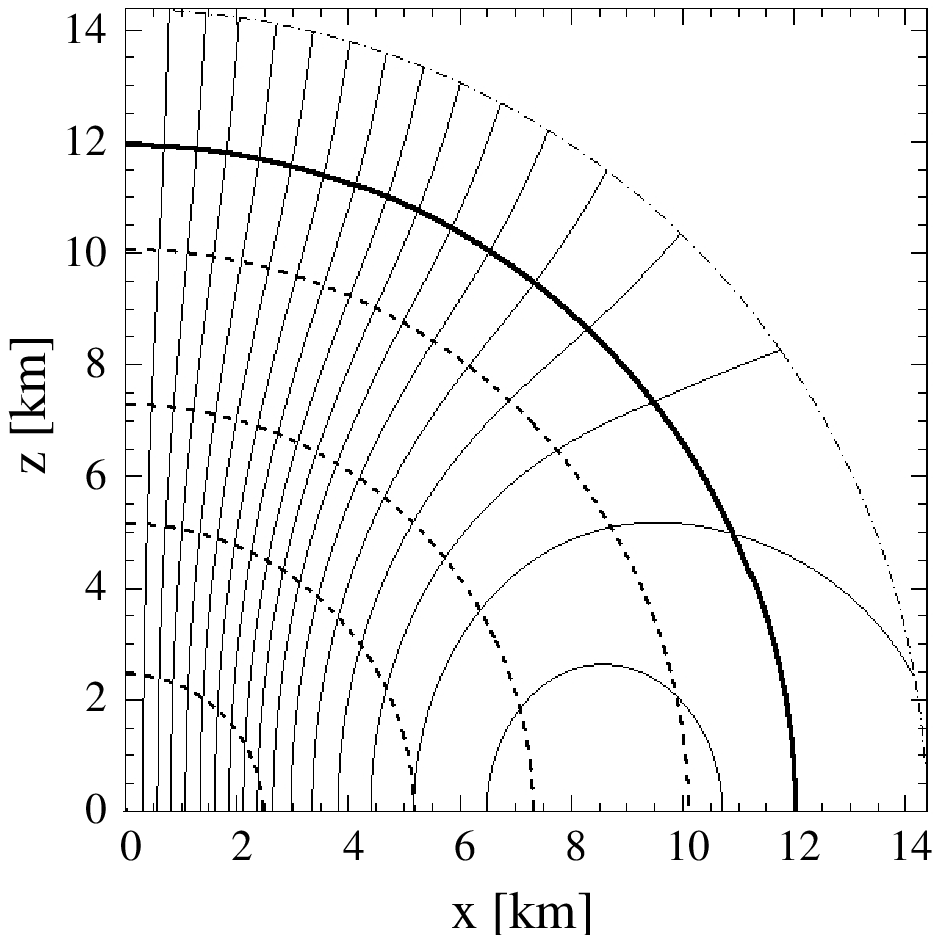}} 
  \caption{Magnetic field lines structure of model MNS3. The thin
           solid lines represent the magnetic field lines while the
           thick dashed lines are rest mass density isocontours for
           $1$, $3$, $5$ and $7 \times 10^{14}$~g cm$^{-3}$. 
           Moreover,
           the thick solid line represents the surface of the
           star, and the thin dash-dotted line marks the boundary 
           of the numerical grid.}
  \label{fig:NSlines}
\end{figure}

As initial models for the magnetized neutron star test, we use the
relativistic self-consistent equilibrium models of \cite{bocquet95},
where all effects of the magnetic field (Lorentz force, spacetime
curvature generated by the magnetic contribution to the
energy-momentum tensor) are taken into account. The equilibrium models
are computed using the LORENE library
\footnote{http://www.lorene.obspm.fr/}. 
We construct non-rotating polytropic equilibrium models with $\Gamma =
2$ and $K=1.455\times 10^{5}$ (cgs units). The central enthalpy is
chosen to be $\ln h_{\rm c}=0.228$, and the magnetic field is that of
a perfect conductor with the current density of \cite{bocquet95} and
vacuum outside.  By increasing the value of the central current
density $j_0$ from $0$ to $5\times10^{14}$\,A\,m$^{-2}$, we compute a
sequence of equilibrium models with a magnetic field ranging from zero
to $1.8\times 10^{16}\,\sqrt{4\pi}$\,Gauss (Table \ref{tab:MNS}).
The magnetic field topology is shown in Fig.\,\ref{fig:NSlines} for a
representative model (MNS3).  It is purely poloidal with field lines
crossing the surface of the neutron star (thick dash-dotted line).  At
sufficiently large distances from the star, the magnetic field has a
dipole topology.

First, we perform simulations in the Cowling approximation, where the
spacetime is kept fixed. We stop the evolution after $5$\,ms which
corresponds to $52\,t_{\rm dyn}$, where $t_{\rm dyn} = \sqrt{r_{\rm e}^3 / M}$
is the characteristic dynamic time-scale of the system. Using the Cowling
approximation, allows us to test the behavior of our MHD scheme without
including yet the coupled evolution of the spacetime itself. The
spacetime fields are computed using the CFC equations in the first
time step, and their values are kept fixed afterwards.  To
carry out convergence tests, we performed computations with models MNS0
and MNS3 on equidistant grids ($n_r \times n_{\theta}$) with
$80\times 10$, $160\times 20$, and $320 \times 40$ zones,
respectively.  The other two models, MNS1 and MNS2, were
simulated only with $160\times 20$ zones. We use the PHM reconstruction
scheme and the KT flux formula in all computations reported in this
section. The neutron star is surrounded by an atmosphere as described
in Sect.\,\ref{sec:vacuum} with a threshold value of $\rho_{\rm thr} =
10^{-7}\, \rho_{\rm max}$, and a floor value $\rho_{\rm atm} =
10^{-9}\, \rho_{\rm max}$. In the highly magnetized models MNS2 and
MNS3 the value of $P_{\rm mag} / P$ is close to the critical value for
the recovery procedure in the outermost zone of the neutron star. In
these models, we raise the threshold value to $\rho_{\rm thr} = 10^{-6}
\, \rho_{\rm max}$ keeping the same floor value.

\begin{figure*}[t!]
  \centering
  \resizebox{!}{0.28\textwidth}{\includegraphics*{0086f5a.eps}} \quad
  \resizebox{!}{0.28\textwidth}{\includegraphics*{0086f5b.eps}}
  \\ [0.5 em] 
  \resizebox{!}{0.28\textwidth}{\includegraphics*{0086f5c.eps}} \quad
  \resizebox{!}{0.28\textwidth}{\includegraphics*{0086f5d.eps}}
  \caption{Evolution of equilibrium configurations of neutron stars in
          the Cowling approximation (fixed spacetime).  The upper
          panels show the central density normalized to its initial
          value for the non-magnetized neutron star MNS0 (left) and the
          magnetized model MNS3 (right). The results are displayed for
          three different grid resolutions ($n_r \times n_{\theta}$):
          $80\times 10$ (dotted), $160\times 20$ (dashed), and
          $320\times 40$ (solid). The lower left panel shows the
          evolution of $\rho/\rho_{\rm c, 0}$ for all computed models for a
          grid resolution of $160 \times 20$: MNS0 (solid), MNS1
          (dashed), MNS2 (dotted) and MNS3 (dash-dotted). The lower
          right panel gives the corresponding Fourier transforms. }
\label{fig:MNSCowling}
\end{figure*}

\begin{figure*}[t!]
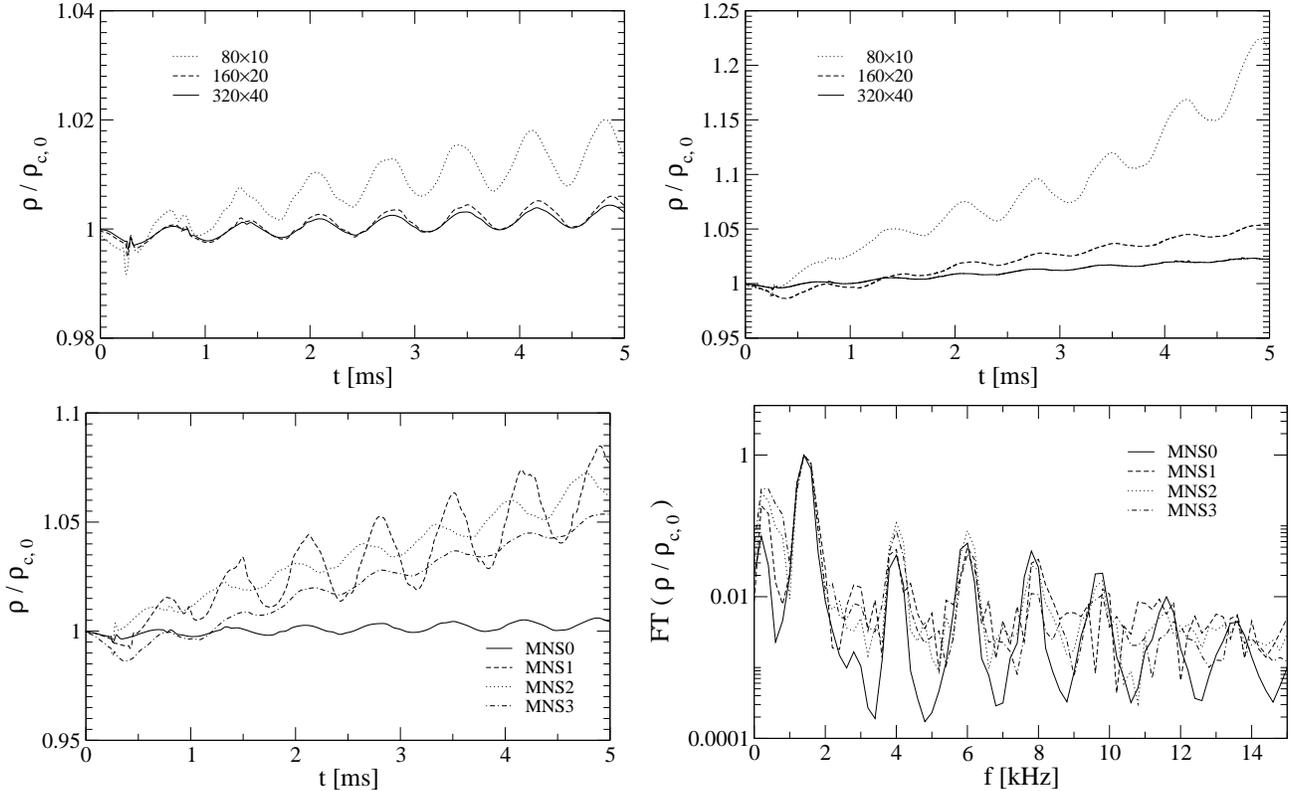

  \centering
  \resizebox{!}{0.28\textwidth}{\includegraphics*{0086f6a.eps}} \quad
  \resizebox{!}{0.28\textwidth}{\includegraphics*{0086f6b.eps}} 
  \\ [0.5 em] 
  \resizebox{!}{0.28\textwidth}{\includegraphics*{0086f6c.eps}} \quad
  \resizebox{!}{0.28\textwidth}{\includegraphics*{0086f6d.eps}}
  \caption{Same as Fig.\,\ref{fig:MNSCowling}, but for a dynamic
           spacetime.}
  \label{fig:MNSDynamic}
\end{figure*}

Three of the panels of Fig.\,\ref{fig:MNSCowling} show the evolution
of the central density with time. Due to numerical truncation errors
in the remapping of the equilibrium model from the
spectral grid used by LORENE to our finite-difference grid, some small
amplitude perturbations are triggered, which excite the normal modes
of pulsation of the star. This causes the periodic oscillations of the
central density. The neutron star remains in equilibrium throughout
its evolution, only a small drift with time is visible in the
central density evolution.  As we increase the grid resolution, this
drift tends to zero.  Comparing the non-magnetized model (MNS0) with
the magnetized models (MNS1, MNS2, and MNS3), the drift,
although small ($<0.2\%$ in the $160\times 20$ models), is
larger in the magnetized models (lower left panel of
Fig.\,\ref{fig:MNSCowling}).  We find that the drift is very sensitive
to the value of $\rho_{\rm thr}$ in model MNS0, if $\rho_{\rm thr} >
10^{-5}\, \rho_{\rm max}$ (for smaller values, there is no
influence). We suppose that, for denser atmospheres there, is a coupling
between the star and the atmosphere that allows a transfer of mass
and momentum from the interior to the atmosphere (see the related
discussion in \citet{stergioulas04}, and \citet{dimmelmeier_06}). This
has two consequences for the evolution: first, the oscillations are
damped more quickly, and second, the slope of the drift changes, even
becoming negative. In the magnetized case, even if the magnetic field
is weak, we have an extra coupling of the interior with the atmosphere
due to the magnetic field lines leaving the star's surface. This
causes an additional very small transfer of mass and momentum from the
atmosphere to the neutron star, which increases the drift in the
evolution (see Fig.\,\ref{fig:MNSCowling}).

The convergence tests show that the order of convergence is $2.13$ and
$1.56$ for model MNS0 and MNS3, respectively. This global order of
convergence is consistent with the second-order accuracy of our
numerical TVD scheme, which reduces to first order at local extrema
such as the center of the star and its surface.

We also compute the Fourier transform of the central density
evolution to obtain the mode frequencies of the neutron star
pulsations (lower right panel of Fig.~\ref{fig:MNSCowling}).
We find the fundamental mode frequency at about $f=2.7$\,kHz,
and subsequent harmonics at $4.6$, $6.4$, $8.2$, $10.0$, $11.8$, and
$13.8$\,kHz, respectively. Since the energy of the magnetic field is
small compared with the potential energy of the star, the influence of
the magnetic field on the mode frequency is small. We find no
frequency difference between the neutron star models within the
frequency resolution ($\sim 0.5$~kHz). We further observe that the
quality of the spectrum deteriorates at higher frequencies for models
with stronger magnetic fields . We suspect that this degradation is an
artifact due to the stronger coupling of the interior with the
atmosphere in the magnetized case.

The second part of the test consists of the evolution of the same
neutron star equilibrium models in a dynamic spacetime. For reasons of
computational efficiency, the CFC equations are computed only every
100th time step, the metric being interpolated in-between as described
by \cite{dimmelmeier_02_a}. The results (Fig.\,\ref{fig:MNSDynamic})
are qualitatively the same as those of the Cowling case discussed
before. The dynamic spacetime causes larger perturbations in the
central density evolution, which now also exhibits a larger drift with
time ($<10\%$ for the $160\times 20$ models). Similar drifts 
were already observed in fully coupled simulations of non-magnetized
\citep{font_02_a} and magnetized models \citep{giacomazzo07}. In both
models, MNS0 and MNS3, the drift reduces with increasing resolution,
and the order of convergence is $3.1$ and $2.5$ respectively. The
convergence order is higher than expected (second order).  We suspect
that this is because the $80\times 10$ zone model is
poorly resolved, \ie the accuracy tends to grow faster than the
order of convergence when doubling the resolution. Regarding the
comparison between magnetized and non-magnetized models (lower left
panel of Fig\,\ref{fig:MNSDynamic}), we observe larger drifts in the
magnetized case due to the stronger coupling with the atmosphere.

The Fourier transform of the central density for the $160\times 20$
models with dynamic spacetime evolution (lower right panel of
Fig.\,\ref{fig:MNSDynamic}) gives a fundamental frequency of
$f=1.4$\,kHz, and higher harmonics at $4.0$, $6.0$, $7.8$, $9.8$,
$11.6$ and $13.7$\,kHz, respectively. We find no dependence on the
amount of magnetization within the frequency resolution.  A similar
result was obtained in the simulations of \citet{montero07}
regarding pulsating and magnetized thick accretion tori around
Schwarzschild and Kerr black holes.  This is unsurprising since the
normal modes of a star are basically sound waves propagating in the
radial direction, and the speed of sound is hardly altered by the
magnetization of the investigated models. However, in the magnetized
case, new modes can appear due to the richer eigenvalue structure of
the GRMHD equations. In particular, it is important to note that
Alfv\'en modes can be excited in the star. For the magnetic field
strengths present in our models, these mode frequencies lie below
$100$\,Hz, \ie much longer simulations are required to be able to
see them in the spectrum. A deeper study of the Alfv\'en modes
performed with our numerical code can be found in \cite{cerda08}.

If we compare the frequencies with those in the Cowling approximation,
we observe that the Cowling approximation tends to overestimate the
frequency of the modes (by almost a factor 2 for the fundamental
mode).  The higher the order of the harmonics, the smaller is the
overestimation, a trend that was observed before in numerical
simulations of purely hydrodynamic models \citep{font_02_a}. The
reason for this behavior is that perturbations on time scales smaller
than the typical time scale of variations in the gravitational field
(which is roughly $t_{\rm dyn}$) behave similarly as in a fixed
spacetime. The frequency corresponding to the dynamic time scale is
$f_{\rm dyn} = 10.4$~kHz. Therefore, modes of frequency higher
than $f_{\rm dyn}$ will be unaffected if the computation is carried out in
the Cowling approximation. This agrees with our mode computations.

\subsection{Core collapse}

\begin{figure}[t!]
  \centering
  \resizebox{!}{0.28\textwidth}{\includegraphics*{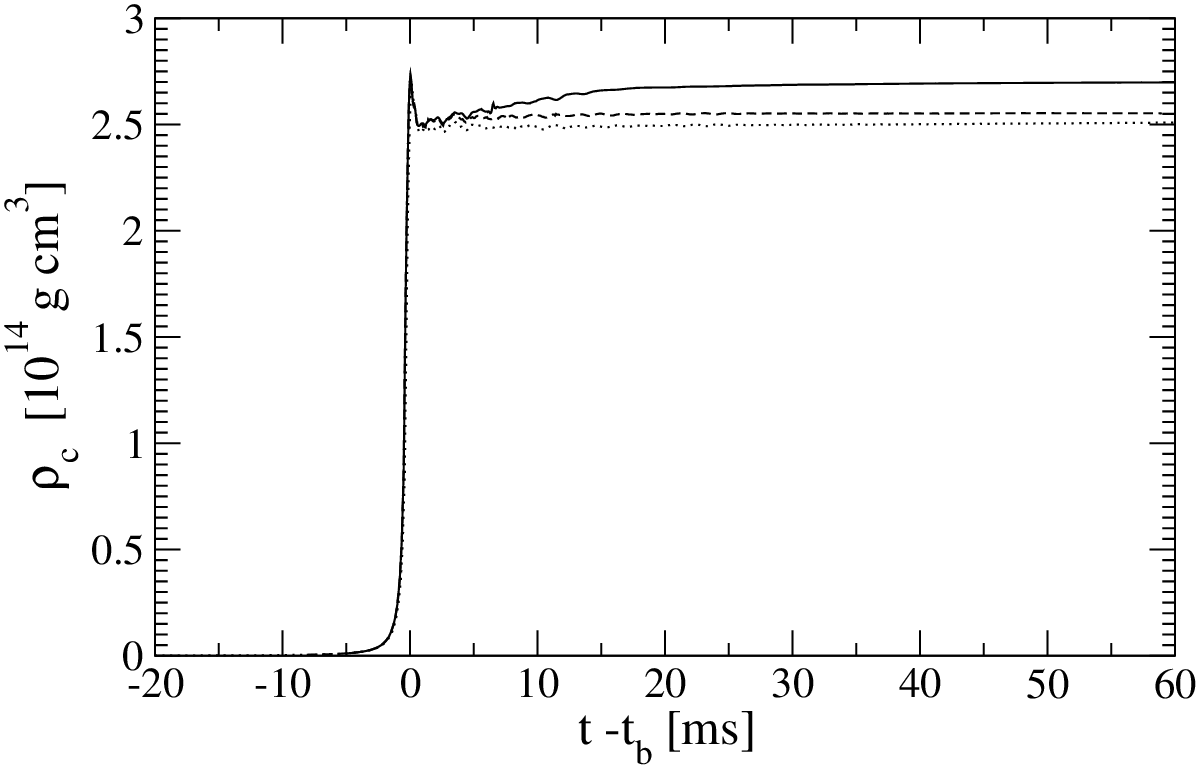}} \quad
  \resizebox{!}{0.28\textwidth}{\includegraphics*{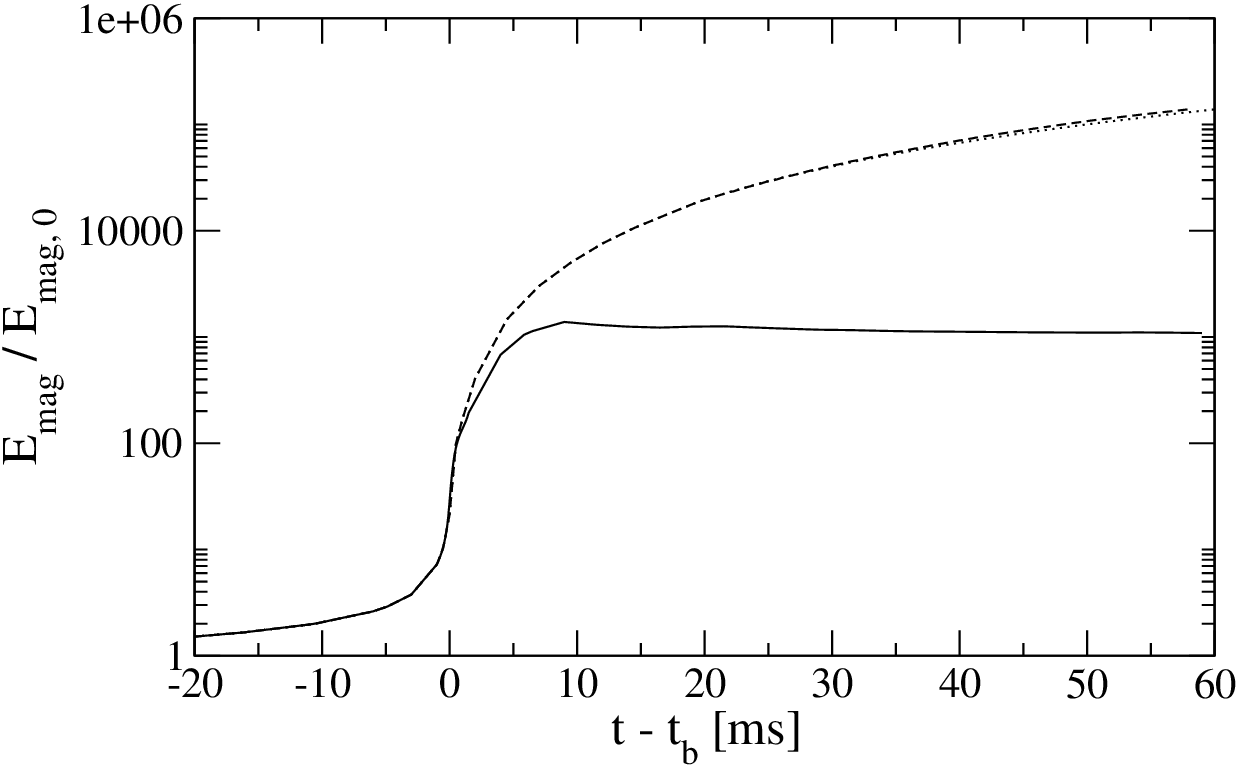}} \quad
  \resizebox{!}{0.28\textwidth}{\includegraphics*{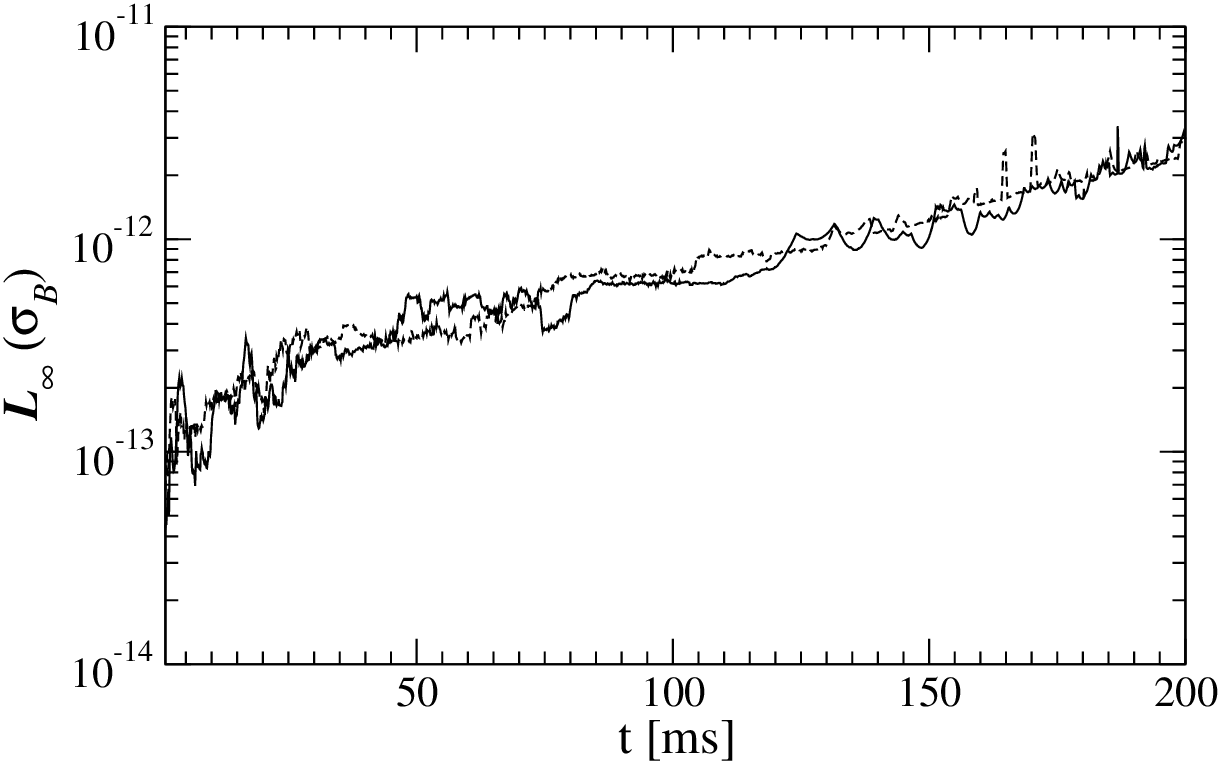}} 
  \caption{ Evolution of the central density $\rho_{\rm c}$ (upper
    panel), of the amplification of the magnetic energy $E_{\rm
      mag}/E_{\rm mag, 0}$ (middle panel) and the $L_\infty$ norm
    of $\sigma_{\rm B}$ (lower panel) for model s20A1B5-D3M12
    (solid), and s20A1B5-D3M10 (dashed), respectively. The results
    obtained with the passive field approximation (model s20A1B5-D3M0
    of \cite{cerda07}) are shown with a dotted line  (almost
      overlapped to the dashed line in the middle panel).
}
  \label{fig:collapse}
\end{figure}

The final test of our numerical code concerns simulations of
magneto-rotational core-collapse. We note that these simulations are
not intended to be of astrophysical relevance, since the treatment 
of neutrinos in the code is still too poor for a study of the
supernova explosion mechanism. Nevertheless, the tests allow us to
validate the code in a fully dynamic context including strong magnetic
fields, realistic stellar progenitors, and a microphysical EOS. To the
best of our knowledge such demanding simulations have not yet been
performed, which highlights the unique potential of our new
numerical code for the study of relativistic stellar core collapse.

\begin{figure*}
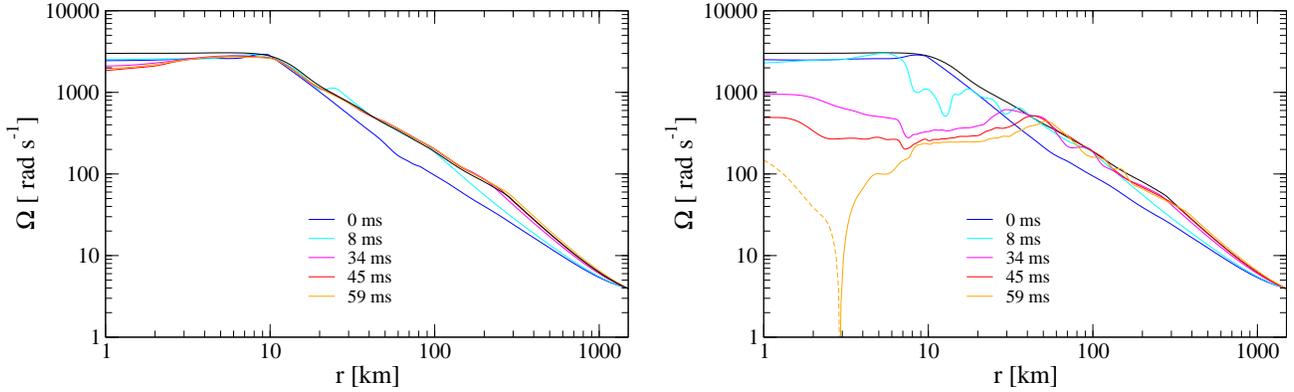

  \centering
  \resizebox{!}{0.28\textwidth}{\includegraphics*{0086f8a.eps}} \quad
  \resizebox{!}{0.28\textwidth}{\includegraphics*{0086f8b.eps}} 
  \caption{Radial profiles of the angular velocity $\Omega$ at the
    equator for model s20A1B5-D3M10 (left panel) and s20A1B5-D3M12
    (right panel) at different times after bounce: $t - t_{\rm b} =
    0$, $8$, $34$, $45$ and $59$\,ms. The time-independent rotation
    profile of the passive field model s20A1B5-D3M0 is shown by the
    black line in both panels. The yellow dashed line in the right
    panel indicates a change of sign of the angular velocity. }
  \label{fig:omegacollapse}
\end{figure*}

As an initial model, we employ the inner part of the iron core of the
solar-metallicity $20\,M_{\odot}$ progenitor model of
\cite{woosley_02_a}.  To this spherically symmetric and
non-magnetized model, we add a rotation profile and a poloidal
magnetic field. The rotation law for the specific angular momentum
is given by $j=A^2 (\Omega_{\rm c} - \Omega)$, where $A=5\times
10^{4}$~km and $\Omega$ is the angular velocity, which has a value
$\Omega_{\rm c}=4.035$~s$^{-1}$ at center. The magnetic field is
generated by a circular current loop of radius $400$~km. This
corresponds to model s20A1B5-D3 in \cite{cerda07} where a more detailed
description can be found. We perform simulations for two
different initial magnetic field strengths, namely for the weakly
magnetized model s20A1B5-D3M10 with a central magnetic field of
$|B|_{\rm c} = 10^{10}\, \sqrt{4\pi}\,$Gauss, and for the strongly
magnetized model s20A1B5-D3M12 with $|B|_{\rm c}=10^{12}\,
\sqrt{4\pi}\,$Gauss. The models are evolved with the tabulated EOS of
\citet{shen_98_a} and an approximate deleptonization scheme
\citep{liebendoerfer_05_a} as described by \citet{dimmelmeier_07_a},
and \citet{cerda07}.  We compare the evolutions of these two models
with that of the corresponding model s20A1B5-D3M0 of \citet{cerda07},
which was evolved with the passive field approximation. Since the
effect of the magnetic field on the collapse dynamics is neglected for
model s20A1B5-D3M0, its evolution should be similar to that of our
weakly magnetized model s20A1B5-D3M10.  The comparison with the
passive field model also allows us to identify genuine MHD effects.

Figure~\ref{fig:collapse} shows the evolution of the central density
(left panel) and the amplification of the magnetic energy (right
panel) for all three models. The latter quantity is computed from
\begin{eqnarray}	
E_\mathrm{mag} & = & \frac{1}{2} \int \mathrm{d}^3 \mb{x} \,
                     \sqrt{\gamma} \, W b^2.
\end{eqnarray}
Additionally the lower panel of Fig.\,\ref{fig:collapse}
shows the $L_\infty$ norm of $\sigma_{\rm B}$ defined as the ratio of the total
magnetic flux at the surface of each numerical cell to the average magnetic
flux on the surface. This dimensionless quantity measures the quality
of the numerical preservation of the divergence of the magnetic field
along the evolution. The final value is consistent with the round-off
error in the evolution, which can be computed as 
$(\rm double\,precision\,accuracy) \times
\sqrt{({\rm number\,of\,iterations})}$ $= 10^{-15} \times \sqrt{5\times 10^{6}}
=2.3\times 10^{-12}$
, if one considers a binomial distribution of errors.
As the collapse proceeds both the density and the magnetic energy grow
very similarly in all three models, because even in the highly
magnetized progenitor model s20A1B5-D3M12 the strength of the magnetic
field is insufficient to affect the collapse dynamics. The ratio
of magnetic energy to gravitational binding energy
\citep[see][]{cerda07} does not exceed a value of $10^{-7}$
($10^{-3}$) during the collapse in model s20A1B5-D3M10
(s20A1B5-D3M12), which justifies the use of the passive field
approximation in the weak magnetic field limit. After core bounce, the
low magnetized model s20A1B5-D3M10 continues to behave similarly
to model s20A1B5-D3M0, since the magnetic field remains weak. The
central density is slightly higher than in model s20A1B5-D3M0,
but the magnetic field is far from saturation and is still
growing linearly with time at the end of the simulation.

On the other hand, the highly magnetized model s20A1B5-D3M12 clearly
shows a saturation of the magnetic field energy shortly after core
bounce. At this time the ratio of magnetic energy to
gravitational binding energy is $7\%$, a value that is never exceeded
during the evolution. Its central density
continues to grow beyond bounce, and the model eventually approaches
an equilibrium configuration with a central density about $10 \%$
larger than in the passive field case.  The behavior of the central
density can be understood by examining the angular velocity profiles
in Fig.\,\ref{fig:omegacollapse}. At the time of bounce, the angular
velocity profile is very similar for all models, since the magnetic
field is still unimportant for the dynamics: the innermost $10$\,km of
the core rotate rigidly, while further out $\Omega$ follows a power
law with an exponent $\sim -1.2$. This profile remains unaltered
during the subsequent evolution of the passive field model. In the
magnetized models, however, the central region spins down, and the
central density rises, the effect being more prominent in the stronger
magnetized model s20A1B5-D3M12. The right panel of
Fig.\,\ref{fig:omegacollapse} shows that the angular velocity begins
to decrease for $10\,{\rm km} \le r \le 30$\,km shortly after
bounce. In this region, the magnetic field is strongest since differential
rotation winds up the magnetic field more efficiently
\citep{cerda07}. On a time scale of about $50$\,ms, the angular
velocity decreases by about a factor $10$, and the innermost few
kilometers of the core even acquire retrograde rotation. 
The reason for this effect is the increasing
magnetic tension in the wound-up magnetic field lines. The
characteristic time scale in which this magnetic tension acts on the
fluid is related to the Alfv\'en crossing time scale of the innermost
region $\tau_{\rm A} \sim 50 $ms, which coincides with the time it takes
for the retrograde rotation to appear.
This effect
was already observed in Newtonian simulations by \cite{mh79}, and
\cite{obergaulinger_06_b}.
For model s20A1B5-D3M10, the spin-down
occurs more slowly, and saturates about $50$\,ms after bounce.

\begin{figure}[t!]
  \centering
  \resizebox{!}{0.28\textwidth}{\includegraphics*{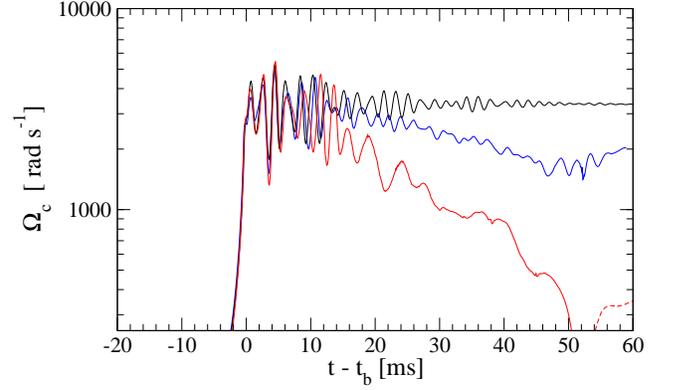}} 
  \caption{Evolution of the central angular velocity $\Omega_{\rm c}$
           for model s20A1B5-D3M12 (red), s20A1B5-D3M10 (blue), and
           model s20A1B5-D3M0 (black), respectively. The red dashed
           line (model s20A1B5-D3M12) indicates a change of sign of
           the angular velocity.}
  \label{fig:omegavst}
\end{figure}

To demonstrate the spin-down more clearly, we plot, in
Fig.\,\ref{fig:omegavst}, the evolution of the central angular velocity
for all three models. In the passive field approximation (black line),
$\Omega$ oscillates after bounce in accordance with the
oscillations of the core, and approaches a constant value at the end of the
simulation. As the magnetic field increases in the progenitor, 
the spin-down of the core occurs more rapidly.
This may be understood by means
of the magneto-rotational instability (MRI hereafter). The MRI is a
shear instability that can appear when both magnetic fields and
differential rotation are present \citep{balbus91}, and it gives rise
to transport of angular momentum.  A necessary condition for the
occurrence of the MRI is $\varpi\partial_{\varpi} \Omega^2 < 0$, where
$\varpi = r \sin{\theta}$.  In unstable regions, the MRI grows
exponentially for all length scales larger than a critical
length-scale $\lambda_{\rm crit} \sim 2 \pi c_{\rm A} / \Omega$, where
$c_{\rm A}$ is the Alfv\'en speed.  The fastest-growing MRI mode
develops on length-scales near $\lambda_{\rm crit}$ on a typical
time-scale of $\tau_{\rm MRI} = 4 \pi [\varpi\partial_{\varpi}
\Omega]^{-1} $.  Therefore, in order to numerically capture the MRI,
one has to resolve length-scales of about $\lambda_{\rm
crit}$. Once the MRI grows, it develops {\it channel flows}
\citep{Hawley1992}, which are unstable to non-axisymmetric
instabilities \citep{Goodman1994} and eventually become turbulent in
three-dimensional simulations \citep{Hawley1996}.

In our simulations, the region with $r > 10$\,km is unstable to the MRI
due to its negative angular velocity gradient. The growth time of the
fastest-growing mode is in the range 1\,ms to 10\,ms for the region
behind the shock wave, and about $1$\,s or even larger further
outside. Since the time scale is independent of the initial magnetic
field strength, these values are similar for both magnetizations
(s20A1B5-D3M10 and s20A1B5-D3M12), and for the passive field case
(s20A1B5-D3M0). However, the critical length scale depends on the
strength of the magnetic field.

\begin{figure}[t!]
  \centering
  \resizebox{!}{0.29\textwidth}{\includegraphics*{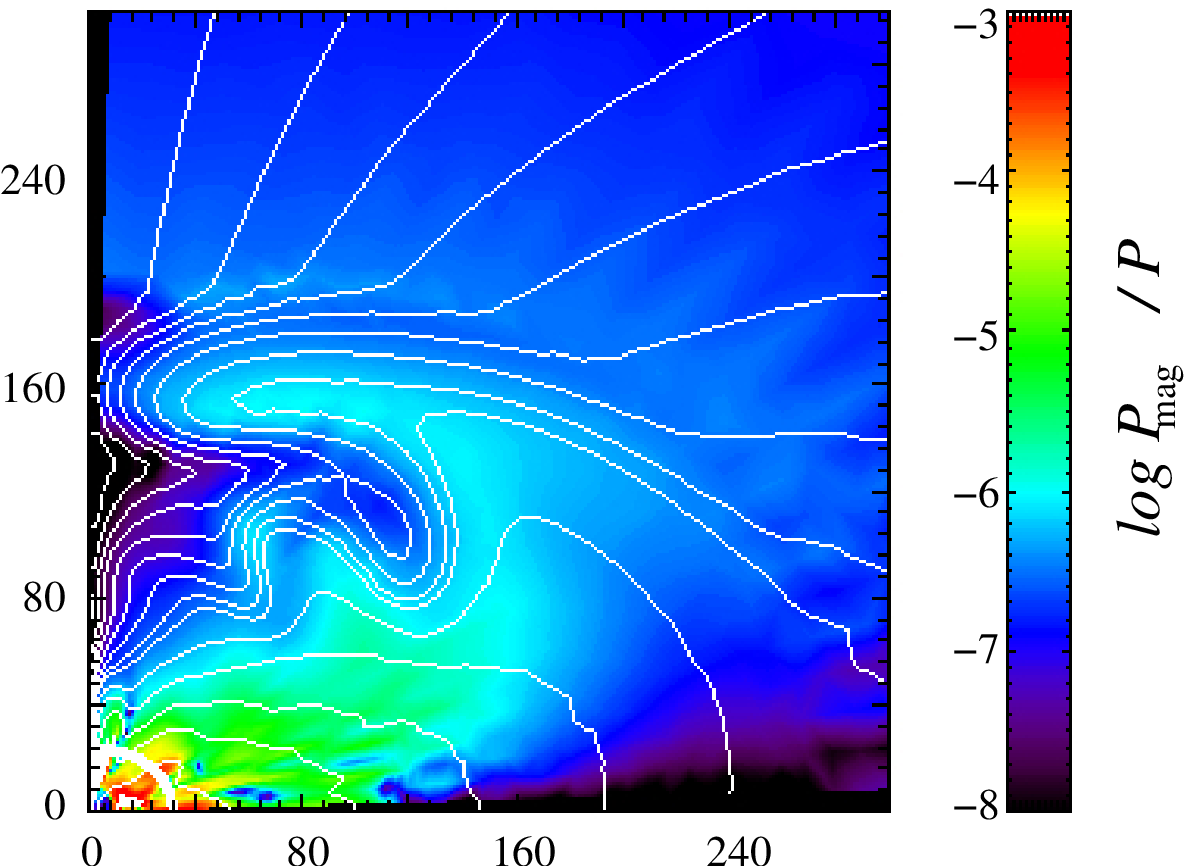}} \quad
  \resizebox{!}{0.29\textwidth}{\includegraphics*{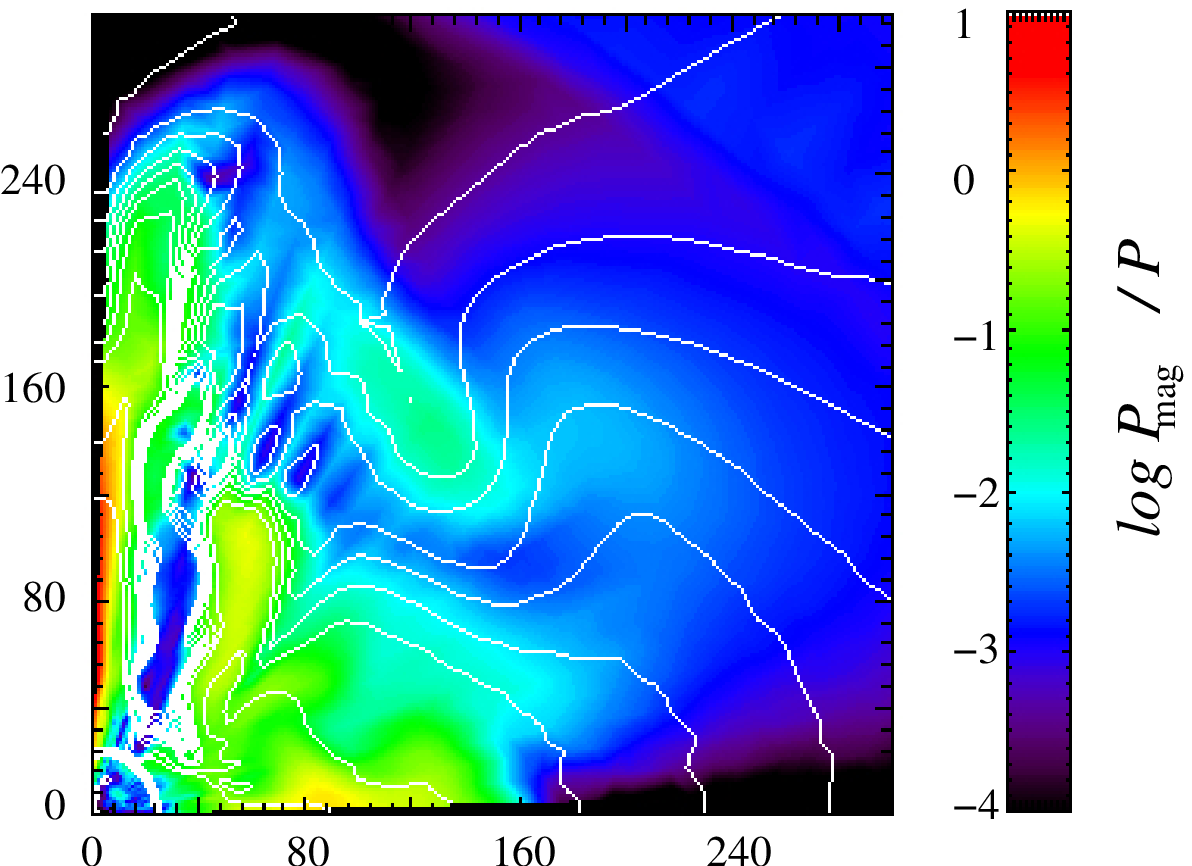}}
  \caption{Magnetic field topology at the end of the simulation,
	   $51$\,ms after bounce, for model s20A1B5-D3M10 (upper
	   panel) and s20A1B5-D3M12 (bottom panel), respectively.  The
	   ratio of magnetic to thermal pressure $P_{\rm mag}/P$ is
	   shown color-coded. Thin, white lines are poloidal magnetic
	   field lines, and the thick, white line marks the
	   neutrino-sphere. The axis labels are in
	   units of km.}
  \label{fig:color}
\end{figure}

For model s20A1B5-D3M12, the critical length scale at bounce is
between $\lambda_{\rm crit}\sim 1\,{\rm km}$ and  $5$\,km inside the unstable
region ($10$\,km $\le r \le 30$\,km).  This region is covered with
$60$ radial and $30$ angular zones, which corresponds to a resolution
($\Delta r, r \Delta \theta$) of $125\,{\rm m} \times 500$\,m at $r
=10$\,km, and $900\,{\rm m} \times 1500$\,m at $r = 30$\,km. This
resolution is marginally sufficient to resolve the length scale of the
fastest-growing mode of the MRI at bounce ($5-10$ radial zones, and
$2-3$ angular zones).  The strong redistribution of the angular
momentum observed for model s20A1B5-D3M12 might therefore be caused by the
MRI. In turn, the saturation of the magnetic field is a direct
consequence of this redistribution of the angular momentum. Without
differential rotation, the poloidal magnetic field cannot be wound up
into a toroidal magnetic field. The typical spin-down time scale
$\tau_{\rm spin-down}$ can be measured from Fig.\,\ref{fig:omegavst}
by fitting an exponential to the declining part of the curve. For
model s20A1B5-D3M12, one obtains $\tau_{\rm spin-down} = 22.5$\,ms,
which corresponds roughly to the time scale of the MRI.

\begin{figure*}[t!]
  \centering
  \resizebox{\textwidth}{!}{\includegraphics*{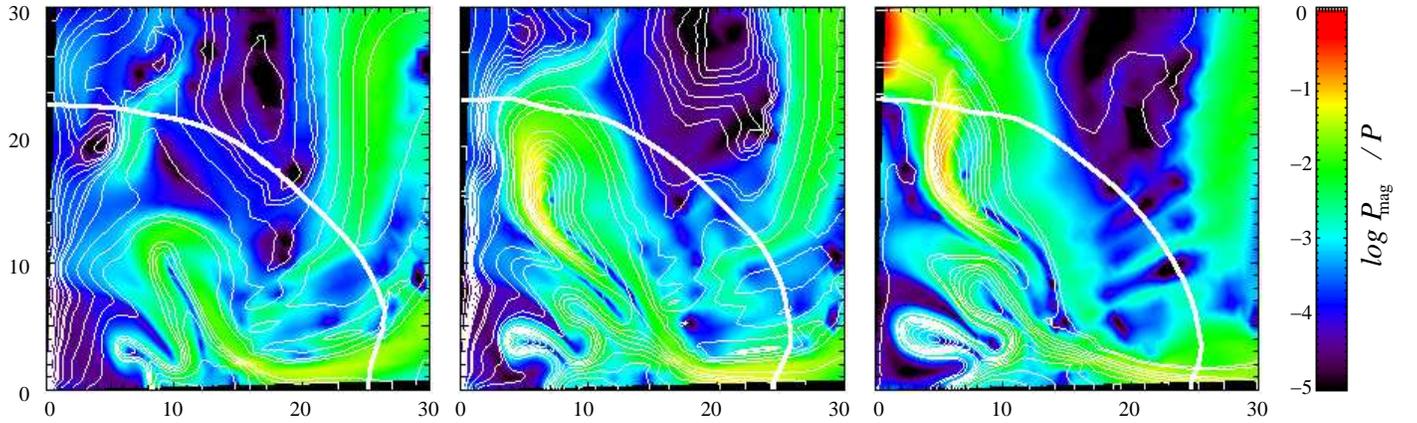}}
  \caption{Details of the magnetic field structure of the core at
	   three different times after bounce: $t-t_{\rm b} = 9$\,ms
	   (left), $11.5$\,ms (middle), and $14$\,ms (right),
	   respectively.  The ratio of magnetic pressure to thermal
	   pressure $P_{\rm mag}/P$ is shown color-coded. Thin, white
	   lines are poloidal magnetic field lines, while the thick,
	   white line marks the neutrino-sphere. The axis
	   labels are in units of km.}  
  \label{fig:colormri}
\end{figure*}

On the other hand, for model s20A1B5-D3M10 the critical length-scale
at bounce is about a factor of $100$ shorter, \ie between $\lambda_{\rm
crit}\sim 10\,{\rm m}$ and $50$\,m, and thus the fastest-growing mode
of the MRI cannot be resolved with our grid resolution. Only modes
with slower growth rates can be resolved on the grid. Accordingly, the
spin-down for this model occurs on a longer time scale of $\tau_{\rm
spin-down} = 62.9$\,ms. At about $50$\,ms after bounce, the innermost
$10$\,km of the core develops a positive angular velocity gradient (see
Fig.\,\ref{fig:omegacollapse}), and hence becomes stable to the MRI.
The central core is no longer able to lose angular momentum, and its
spin down stops.
We suspect that the appearance of this positive gradient
is due to the poorly resolved MRI, which turns out to be more 
efficient in the inner region, where the resolution is higher, instead
of where the shear is larger.
The magnetic field continues to grow at similar rates
until the end of the simulation due to the further winding-up of
poloidal magnetic field lines, and because angular momentum transport
is insufficient to affect the rotation profile outside the innermost
$10$\,km significantly.

Figure~\ref{fig:color} displays the magnetic field topology for models
s20A1B5-D3M10 and s20A1B5-D3M12 at the end of the simulation. The low
magnetized model s20A1B5-D3M10 (top panel) has a similar field
structure as model s20A1B5-D3M0 of \cite{cerda07}, since the passive
field approximation holds very well for weakly magnetized progenitors
(apart from its inability to capture the MRI). The prompt convection
\footnote{This transient is produced by an unstable entropy gradient,
which is probably an artifact of our poor neutrino treatment.  The
interested reader is addressed to \cite{cerda07} for a detailed
discussion of this issue.}
developing after bounce twists the magnetic field outside the
neutrino-sphere, which is assumed to be located at $\rho_{\nu} =
2\times 10^{12}$\,g\,cm$^{-3}$, at about $30$~km.
In model s20A1B5-D3M12, the magnetic field grows to values close to
equipartition, and a distinctive, strongly magnetized outflow
propagates along the axis behind the shock front. Between $10\,{\rm
km} \la r \la 30$\,km, where the MRI is predominantly growing,
axisymmetric channel flows form, which are morphologically similar to
the flows found in the simulations of \cite{Hawley1992}. We analyze
this issue in more detail in Fig.\,\ref{fig:colormri}, where the
development of the channel flows is shown. Their length scale
increases as the magnetic field becomes stronger, and since we assume
axisymmetry they are stable, \ie they do not cause any turbulence.

Another important difference between models s20A1B5-D3M10 and
s20A1B5-D3M12 is the location of the shock. At $\sim 50$\,ms
after core bounce, the shock is located about $50$\,km further out
in the strongly magnetized model s20A1B5-D3M12 than in the weakly
magnetized model s20A1B5-D3M10. This is most likely a consequence of
the transport of angular momentum by the MRI which pushes the shock
front to a larger radial distance, although our current grid
resolution is probably too poor in the shock region to confirm this
interpretation conclusively. Understanding this effect and, in particular, its
implications for the explosion mechanism, requires a separate study,
which will be published elsewhere.

Since no other simulations yet have been published that are capable of treating
a similar combination of general relativity and microphysics, it is
impossible to compare directly with other work.
Nevertheless, we find qualitative agreement with related simulations
of magneto-rotational core collapse \citep{obergaulinger_06_a,
obergaulinger_06_b, shibata_06_a, Burrows07}.  In particular, our
simulations share the following aspects with these investigations: (i)
redistribution and transport of angular momentum radially outwards due
to the MRI, resulting in the spin down of the central region of the
core; (ii) increase of the central density after core bounce due to
angular momentum losses; and (iii) appearance of a weakly
relativistic but highly magnetized outflow along the axis.  This
agreement strengthens our confidence in the suitability of our new
numerical code for the systematic investigation of magneto-rotational
core collapse, which we shall report elsewhere.

\section{Conclusions}
\label{sec:conclusions}
We have presented a new numerical code that solves the GRMHD equations
coupled to the Einstein equations for the evolution of a dynamic
spacetime. Hence, it extends the small list of available codes that
are capable of modeling these challenging physics.  The main objective
of the new code is the study of astrophysical scenarios in which both
strong magnetic fields and strong gravitational fields are present,
such as the magneto-rotational collapse of stellar cores, the
collapsar model of GRBs, and the evolution of neutron stars.

Our new numerical code is based on high-resolution shock-capturing
schemes to solve the flux-conservative hyperbolic GRMHD equations, and
the constraint-transport method to ensure the solenoidal
condition of the magnetic field. The Einstein equations are formulated
in the CFC approximation, and the resulting
elliptic equations are solved using a linear Poisson solver. The
motivation to use CFC is based on the astrophysical applications
envisaged for the code, which do not deviate significantly from
spherical symmetry. Furthermore, the code incorporates several
equations of state, ranging from simple analytical expressions to
tabulated microphysical equations of state. 

We have presented a number of stringent tests of our new GRMHD
numerical code, which are the main focus of this paper. The test
calculations demonstrate the ability of the code to handle properly
all aspects appearing in the astrophysical scenarios the code is
intended for, namely relativistic shocks, strongly magnetized fluids,
and equilibrium configurations of magnetized neutron stars.  One of
the tests the code has passed successfully is in fact an application,
namely the simulation of general relativistic magneto-rotational core
collapse using a realistic stellar progenitor model and a
microphysical equation of state. We have compared the results obtained
by our new code with those of a previous study based on the 
passive magnetic-field approximation, and find good agreement for initially
weakly magnetized progenitors.

Finally, we mention that the new code is also capable of handling the
gravitational collapse leading to the formation of a black hole.
Results for this specific application will be presented
elsewhere. Further extensions of the code that we foresee in the near
future include the incorporation of a simplified scheme for neutrino
transport (to explore the post-bounce evolution of collapsing
magnetized cores more reliably) along with the implementation of
resistive MHD.


\begin{acknowledgements}
This research has been supported by the Spanish {\it Ministerio de
Educaci\'on y Ciencia} (grant AYA2004-08067-C03-01), and by the
Collaborative Research Center on {\it Gravitational Wave Astronomy} of
the Deutsche Forschungsgesellschaft (DFG SFB/Transregio 7).
We would like to thank J. Novak, for the C++ subroutines
to import the magnetized equilibrium models from the Lorene code. 
We also thank the referee, L. Rezzolla, for his useful comments and 
suggestions.
\end{acknowledgements}


\bibliography{0086}

\end{document}